\documentclass[aps,prl,twocolumn, showpacs,superscriptaddress,10pt]{revtex4-2}
\usepackage[USenglish]{babel}

\usepackage{amssymb}   
\usepackage{amsmath}
\usepackage{placeins}
\usepackage[T1]{fontenc} 
\usepackage[latin1]{inputenc}
\usepackage{lmodern} 
\usepackage{color}
\newcommand{\ket}[1]{\left|#1\right\rangle}

\newcommand{\bra}[1]{\left\langle#1\right|}

\usepackage{float}
\usepackage{graphicx,xcolor}
\usepackage{amsmath, bbm}
\usepackage{hyperref}
\usepackage{epstopdf}
\usepackage{bbm}
\usepackage{soul}
\newcommand{\mean}[1]{\langle #1\rangle}

\renewcommand{\imath}[0]{\mathrm{i}}

 \usepackage[normalem]{ulem}
\begin{document}
\title{Coherent electron-vibron interactions in Surface-Enhanced Raman Scattering (SERS)}

\author{Miguel \'A. Mart\'inez-Garc\'ia}
\author{Diego Mart\'in-Cano}
\email{diego.martin.cano@uam.es}
\affiliation{Departamento de F\'{i}s\'{i}ca Te\'{o}rica de la Materia Condensada and Condensed Matter Physics
Center (IFIMAC), Universidad Aut\'{o}noma de Madrid, E28049 Madrid, Spain}

\date{\today}

\begin{abstract}
In this work we identify coherent electron-vibron interactions between near-resonant and non-resonant electronic levels that contribute beyond standard optomechanical models for off-resonant or resonance SERS. By developing an open-system quantum model using first molecular interaction principles, we show how the Raman interference of both resonant and non-resonant contributions can provide several orders of magnitude modifications of the SERS peaks with respect to former optomechanical models and over the fluorescence backgrounds. This cooperative optomechanical mechanism allows for generating an enhancement of nonclassical photon pair correlations between Stokes and anti-Stokes photons, which can be detected by photon-counting measurements. Our results demonstrate Raman enhancements and suppressions of coherent nature that significantly impact the standard estimations of the optomechanical contribution from SERS spectra and their quantum mechanical observable effects.
\end{abstract}

\maketitle

The discovery of enormous enhancements on Raman molecular signals in close proximity to metallic surfaces~\cite{Fleischmann1974}, attributed to surface plasmons and charge-transfer effects~\cite{Jeanmaire1977,Albrecht1977}, led to the first experimental demonstrations of single molecule vibrational spectroscopy~\cite{Kneipp1997,Nie1997} and the opening of SERS research~\cite{Zrimsek2016}. Despite the robust evidence of such observations, several questions and discrepancies have arisen about the  generating mechanisms in SERS and explanations of uncommon spectroscopic features with respect to standard Raman spectroscopy~\cite{Moskovits2013,Zrimsek2016}. 

Motivated by anomalously large intensity peaks in SERS spectra, a recent theoretical work~\cite{Roelli2016} proposed an optomechanical model for SERS that support nonlinear vibrational amplifications in analogy to cavity optomechanics~\cite{Aspelmeyer2014}, where the cavity role is played by localized surface plasmons, and the mechanical oscillations correspond to intramolecular vibrations. Several theoretical works have then explored multiple molecular optomechanical phenomena~\cite{Esteban2022}, including the generation of nonclassical photon correlations~\cite{Schmidt2016,Schmidt2016}, optical-THz frequency conversion~\cite{Roelli2020}, resonance SERS models~\cite{Neuman2019,Neuman2020}, anharmonic regimes~\cite{Dezfouli2019} and memories~\cite{Gurlek2021}. The following experimental studies include pioneer  evidences confirming some of such optomechanical predictions~\cite{Benz2016,Lombardi2018,Chen2021a,Xomalis2021,Chen2021}, but also showing anomalous higher couplings ~\cite{Benz2016,Lombardi2018} with respect to those predicted by standard optomechanical models, and intricate photoluminescence phenomena~\cite{Chen2021a,Chen2021}. In this work we study the coherent optomechanical mechanism in SERS between near-resonant and non-resonant electronic levels, and their influence in modifying the predictions from previous molecular optomechanical models.
\begin{figure}[t!]
	\begin{center}
		\includegraphics[width=8.6cm]{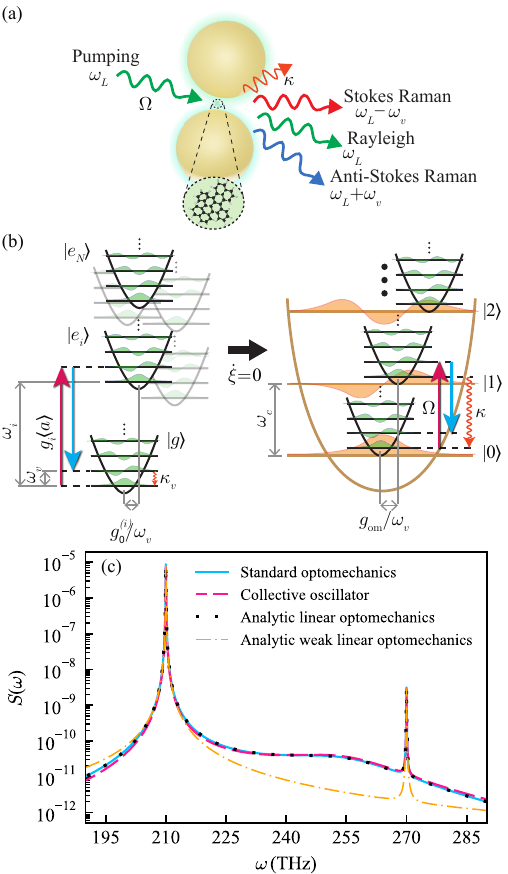}
	\end{center}
		\vspace{-0.2cm}
	\caption{(a) Sketch of the SERS system. (b) Energy level diagram for the displaced oscillator model for several electronic transitions (left) and the cavity optomechanical model as a result of an adiabatic elimination of the electronic degrees of freedom (right). (c) Emission spectra for the  adiabatic elimination Hamiltonian (blue line), the collective oscillator approximation  $H_\mathrm{B}$ (magenta dashed), full spectral analytical solution of a linearized cavity optomechanical Hamiltonian~\cite{Weiss2013a}  (black dotted) and simplified analytical formula without fluctuations~\cite{Wilson-Rae2008} (yellow dash-dot). We consider a molecule with a strong Raman polarizability with $N=1000$,  $g_\mathrm{om}=0.1~\mathrm{THz}$, $\omega_i/2\pi=673.0~\mathrm{THz}$. The rest of the parameters are fixed to  $\omega_c/2\pi=270~\mathrm{THz}$, $\omega_v/2\pi=30~\mathrm{THz}$,  $\Omega/2\pi=4~\mathrm{THz}$, $g/2\pi=2.5~\mathrm{THz}$, $g_0/2\pi=3~\mathrm{THz}$,  $\kappa/2\pi=33~\mathrm{THz}$, $\kappa_v/2\pi=60~\mathrm{GHz}$ and  $\gamma/2\pi=50~\mathrm{MHz}$.}\label{Sketch_Model}
\vspace{-0.6cm}
\end{figure}

Figure~\ref{Sketch_Model}(a) displays a typical sketch of a SERS scenario, in which a laser shines a single organic molecule close to metallic nanoparticles  producing elastically scattered light, fluorescence and Raman signals. The interaction between electrons and vibrations make molecules intrinsic optomechanical systems~\cite{Reitz2020,Gurlek2021}, with strong Raman cross-sections for certain species, such as polycyclic aromatic hydrocarbons~\cite{Negri2002, Zirkelbach2021}. Such electron-vibron interactions are responsible for the generation of Raman signals~\cite{Tommasini2009}, which can be largely enhanced by the extreme confinement of surface plasmons involved in single molecule SERS~\cite{Kneipp1997,Nie1997,Benz2016,Zrimsek2016}.

In order to account both resonant and off-resonant Raman interactions, we consider as a basis of our model $N$ electronic excited levels $\ket{e_i}$ that accept optical transitions from a ground state $\ket{g}$ with frequencies $\omega_i$ and annihilation operator $\hat{\xi}_i=\ket{g}\bra{e_i}$. These transitions are linearly coupled to intramolecular vibrations~\cite{Reitz2020} with  strengths  $g_{0,i}$  , annihilation operator $\hat{v} $ and frequencies $\omega_v$ (c.f. Fig.\ref{Sketch_Model} (b)). This model, also commonly known as the displaced oscillator model~\cite{May2004,Guthmuller2016}, is well known to provide the Franck-Condon spectroscopic principles~\cite{Reitz2020}   and a microscopic connection with the Raman polarizability tensor~\cite{Tommasini2009}, valid for small vibrational amplitudes. To describe the main electromagnetic coupling in SERS,  we consider that such optical electronic transitions are either resonantly or dispersively coupled to a single cavity mode with a  strength $g_i$, corresponding to a plasmon resonance with annihilation operator $\hat{a}$, and driven by a laser with Rabi strength $\Omega$ and frequency $\omega_L$. Therefore, the resulting Hamiltonian in our minimal approach within the rotating frame of laser is the following (see derivations details in Supplemental Material (SM)~\cite{supp3}\nocite{Garraway2011})
\begin{align}
\label{eqn:Hamiltonian}
	H&= \sum^N_{i=1} \hbar \Delta_i \hat{\xi}_i ^\dagger\hat{\xi}_i+ \hbar \Delta_c \hat{a} ^\dagger\hat{a}+ \hbar \omega_v \hat{v}^\dagger \hat{v}+\hbar\Omega(\hat{a}^\dagger + \hat{a})\\  \nonumber
	&+\sum^N_{i=1}\hbar g_{0,i} \hat{\xi}_i^\dagger \hat{\xi}_i (\hat{v}^\dagger +\hat{v})+\sum^N_{i=1} \hbar g_i (\hat{\xi}_i^\dagger \hat{a} +  \hat{\xi}_i \hat{a}^\dagger),
\end{align}
where the laser detunings with respect to the optical transitions and the plasmon resonance are denoted by $\Delta_i=\omega_i-\omega_L$ and  $\Delta_c=\omega_c-\omega_L$, respectively. Moreover, we consider an open-system approach~\cite{Gardiner2004}  where the cavity, vibron and the electronic degrees of freedom couple respectively to different bath continua. For most typical molecular parameters, the system couplings strengths lie far from their bare frequency values and so, their excitations follow a Markovian decay to such baths. We describe this scenario by a standard master equation (SME)~\cite{Gardiner2004}  $\dot{\hat\rho}=-i[\hat H,\hat\rho]+\sum_i\frac{\gamma_i}{2}\mathcal{L}(\hat\xi_i)+\frac{\kappa}{2}\mathcal{L}(\hat a)+\frac{\kappa_v}{2}\mathcal{L}(\hat v)$,
where we have defined the Lindblad superoperator $\mathcal{L}(\hat O)=2\hat O\hat \rho \hat O^\dagger-\hat O^\dagger \hat O\hat \rho-\hat \rho \hat O^\dagger \hat O$, characterizing  dissipative mechanisms for spontaneous emission $\gamma_i$, vibron decay $\kappa_v$ and cavity decay $\kappa$, being the latter the responsible for the emission background in the cavity spectrum for quantum optomechanical models~\cite{Schmidt2021}. The involved parameters of these models can be estimated from ab-initio calculations  \cite{Kumar1992,Guthmuller2016} and experimental measurements. In this work we will account for typical orders-of magnitude estimations from both sources, in close correspondence with those achievable in state-of-the-art nanoparticle on-a-mirror configurations~\cite{Benz2016,Lombardi2018,Chen2021,Chen2021a} and strong Raman molecular scatterers with strong optical transitions as polycyclic aromatic hydrocarbons~\cite{Clear2020,Zirkelbach2022,Zirkelbach2023}.

An important feature of the Hamiltonian (\ref{eqn:Hamiltonian}) is that we can map it into the \textit{standard optomechanical Hamiltonian} for SERS~\cite{Roelli2016}, $H_{\mathrm{om}}=\hbar \Delta^\prime_c \hat{a} ^\dagger\hat{a}+ \hbar\omega_v \hat{v}^\dagger \hat{v}+\hbar\Omega(\hat{a}^\dagger + \hat{a})+\hbar g_{\mathrm{om}} \hat{a}^\dagger \hat{a} (\hat{v}^\dagger + \hat{v})$, via an adiabatic elimination of the electronic degrees of freedom  to first order in driven cavity field, i.e. $\dot{\xi}_i\approx0\rightarrow{\hat{\xi}_i\approx -g_i \hat{a}/\Delta_i}$. This approximation leads to the following relation for the optomechanical coupling strength $g_{\mathrm{om}}$ that is obtained directly from microscopic parameters (see detailed derivation in SM~\cite{supp3})
\begin{align}
	\label{eqn:OptomechanicalCoupling}
	g_{\mathrm{om}}=\sum^N_{i=1} g_{0,i}  \frac{g_i^2}{\Delta_i^2}. 
\end{align}
 Equation~(\ref{eqn:OptomechanicalCoupling}) establishes a direct connection from the fundamental microscopic couplings to both, the optomechanical constant and the Raman polarizability tensor $\alpha$, and thus its dipole moment ($g_{\mathrm{om}}\propto\alpha E$), as performed in the original derivation of the standard quantum optomechanical model of SERS~\cite{Roelli2016,Schmidt2017}. We notice that similar formulas have been derived in the absence of plasmonic environments for closed quantum systems~\cite{Tommasini2009,Guthmuller2016}. We will use the mathematical connection in Eq.~\ref{eqn:OptomechanicalCoupling} with the optomechanical constant evaluated from the Raman activity~\cite{Roelli2016}, along with order-of-magnitude evaluations of the remaining parameters, to estimate the effective number of electronic levels $N$ involved in the Raman tensor. Such estimation does not affect the predictions of the optomechanical Hamiltonian as long as the adiabatic elimination holds, and it can be always substituted by a more realistic evaluation from ab-initio molecular calculations. We also point out that, as a result of the adiabatic elimination, there is a red shift in the cavity frequency, $-\sum^N_{i=1}g_i^2/\Delta_i$, arising from the dielectric response of the molecule, which can be substantial due to the small mode volumes of plasmonic particles and large number of electronic levels. For consistency in the comparison, we will renormalize the cavity frequency to map our  model with the standard molecular optomechanical Hamiltonian, i.e., $\Delta_c \rightarrow{\Delta^\prime_c=\omega_c^\prime-\omega_L}$. 

Despite the larger generality of the Hamiltonian~(\ref{eqn:Hamiltonian}) compared to the standard optomechanical Hamiltonian, the extensive number of degrees of freedom involved in the calculation of Raman spectra \cite{Kumar1992,Guthmuller2016} calls for demanding numerical methods~\cite{delPino2018,Fowler-Wright2022,Cygorek2022}  or approximations to solve it beyond the direct adiabatic elimination. In this line, the similar magnitudes of the electron-vibron interaction strengths and optical transition frequencies allows for considering a relevant simplification for the case of identical couplings and frequencies, i.e. $g_{0,i}=g_0$, $g_i=g$ and $\omega_{i}=\omega_{0}$. This reduction enables us to treat  all the electronic levels approximately as quantum harmonic oscillators~\cite{Kurucz2010}. 

Therefore, the Hamiltonian~(\ref{eqn:Hamiltonian}) can be approximated in terms of a collective bright bosonic operator (with annihilation operator, $\hat{b}= g \sum_i{\xi}_i/\sqrt{N}$),  due to the negligible contributions of the dark electronic states  for large off-resonant drivings, which leads to the following Hamiltonian  (see derivation details  in SM~\cite{supp3})
\begin{align}
	\label{eqn:HamiltonianHolstein}
	H_{\mathrm{B}}&=\hbar \Delta_b \hat{b} ^\dagger\hat{b}+ \hbar \Delta_c \hat{a} ^\dagger\hat{a}+ \hbar \omega_v \hat{v}^\dagger \hat{v}+\hbar\Omega(\hat{a}^\dagger + \hat{a})\\  \nonumber
	&+ \hbar g_0 \hat{b}^\dagger \hat{b} (\hat{v}^\dagger + \hat{v})+\sqrt{N} \hbar g (\hat{b}^\dagger \hat{a} +  \hat{b} \hat{a}^\dagger),
\end{align}
where $\Delta_b=\omega_0-\omega_L$. 

Figure~\ref{Sketch_Model}(c) shows the outcome of the steady-state cavity spectrum performed in Qutip~\cite{Johansson2012}, $S(\omega)=\frac{1}{\pi}\int_0^\infty e^{i(\omega-\omega_L)\tau}\langle\hat a^\dagger(0)\hat a(\tau)\rangle d\tau$, of such model (magenta dashed line), compared to the standard optomechanical Hamiltonian resulting from the adiabatic elimination (blue solid line) for a configuration of cavity cooling \cite{Aspelmeyer2014}  and  typical optomechanical SERS parameters (c.f. caption). The results show an excellent agreement between both models provided by the connection given in formula (\ref{eqn:OptomechanicalCoupling}). For comparison purposes, we have also plotted the full spectral solution of a linearized cavity optomechanical Hamiltonian~\cite{Weiss2013a} (black dotted line) and the commonly-used analytical formula in the absence of fluctuations, valid for intense cavity drivings~\cite{Wilson-Rae2008,Roelli2016}. Whereas the agreement of the former with our results demonstrate the perfectly linearized character of the optomechanical Hamiltonians for  typical molecular  parameters of  largely detuned transitions, we notice that despite the simplified analytical formula (ref.~\cite{Wilson-Rae2008,Roelli2016}) succeeds in describing the coherent narrow anti-Stokes Raman peak resulting from optically induced vibrons, it cannot provide the broader background coming from the incoherent fluctuations of the system \cite{Schmidt2021} that becomes relevant at the moderate drivings achievable in SERS,  i.e., $\mean{\hat{a}^\dagger\hat{a}}	\lesssim 1$. 

\begin{figure}
	\begin{center}
		\includegraphics[width=8.6cm]{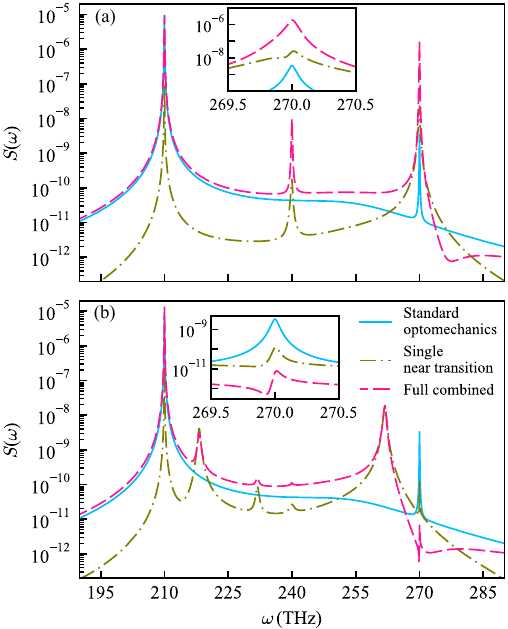}
	\end{center}
	\vspace{-0.2cm}
	\caption{Emission spectra $S(\omega)$ for the optomechanical Hamiltonian $H_\mathrm{om}$~\cite{Roelli2016} (solid blue), the resonant SERS Hamiltonian $H_\mathrm{res}$~\cite{Neuman2019} (dashdot olive green), and the total Hamiltonian $H_{\mathrm{om+ res}}$ containing both contributions (dashed magenta). Panel (a) describes an enhancement scenario of the anti-Stokes peak with the resonant transition at $\omega_r/2\pi=270~\mathrm{THz}$ whereas (b) describes a case of suppression with the ZPL at $\omega_r/2\pi=262~\mathrm{THz}$. The plasmon detuning is set at $\Delta_c^\prime=15~\mathrm{THz}$, whereas $g_\mathrm{om}/2\pi=0.1~\mathrm{THz}$, $g_{0,r}/2\pi=3~\mathrm{THz}$, $g_{r}/2\pi=2.5~\mathrm{THz}$ and $N=1000$. The rest of the parameters coincide with Fig.~\ref{Sketch_Model} (c).} 
	\label{CooperativeRamanNearandOff}
	\vspace{-0.6cm}
\end{figure}
Zero-phonon line (ZPL) transitions that are nearly resonant either with the cavity mode or the laser, are not susceptible to be adiabatically eliminated and thus must be kept at equal footing with the cavity dynamics. Therefore, considering the case of a single near-resonant transition together with the off-resonant ones, the Hamiltonian~\ref{eqn:Hamiltonian} turns to a sum of the adiabatic eliminated Hamiltonian plus the terms associated to the single electronic transition
\begin{align}
	\label{eqn:AdiabaticEliminationHamiltonian}
	H_{\mathrm{om+ res}}&= H_{\mathrm{om}}+\hbar \Delta_r \hat{\xi}_r ^\dagger\hat{\xi}_r+ \hbar g_{0,r} \hat{\xi}_r^\dagger \hat{\xi}_r (\hat{v}^\dagger + \hat{v})\\ \nonumber &+ \hbar g_r (\hat{\xi}_r^\dagger  \hat{a}+ \hat{\xi}_r \hat{a}^\dagger ),
\end{align}
where  $g_{0,r}$ is its electron vibrational coupling, $g_r$ its cavity coupling strength and $\Delta_r$ its frequency detuning. Notice that for consistency, the single near-resonant transition contribution should be removed  in the calculation of the Raman polarizability (c.f. formula~(\ref{eqn:OptomechanicalCoupling})), but this consideration in the total optomechanical coupling is generally negligible due to the large number of off-resonant transitions contributing to it. In Fig.~\ref{CooperativeRamanNearandOff}(a), we represent  the cavity spectrum for this model (dashed magenta line) and compare it with the cases of a purely resonance SERS model~\cite{Neuman2019} (dash-dot olive line) and the off-resonant optomechanical Hamiltonian~\cite{Roelli2016} (solid blue line) for a cavity cooling configuration with an emitter ZPL detuned slightly above (panel a) and below (panel b) the anti-Stokes frequency.
Interestingly, we find several orders of magnitude enhancements (a) and supressions (b) of the anti-Stokes amplitudes compared to the previous resonant SERS~\cite{Neuman2019} and standard optomechanical~\cite{Roelli2016} models. Such modifications cannot be simply explained by a dominance of the near-resonant transition, nor a incoherent sum of both independent spectra, and thus it points out to a coherent behaviour of the Raman scattering between off-resonant and near-resonant transitions. 

To understand such cooperative behavior, we develop analytical formulas of the cavity spectra for our model following similar procedures and approximations as in optomechanics~\cite{Neuman2019}. We consider the cavity field evolution provided approximately by its bare coupling to the electronic transitions, with a separation of their scattering contributions generated  by fluctuations from resonance fluorescence $\delta a_\mathrm{RF}$, Raman $\delta  a_\mathrm{R}$ and hot luminescence $\delta  a_\mathrm{H}$, respectively. The resulting steady-state cavity spectrum  for the Raman fluctuations, $S_\mathrm{R}\propto\int_0^\infty e^{i(\omega-\omega_L)\tau}\langle \delta \hat a_\mathrm{R}^\dagger(0) \delta\hat a_\mathrm{R}(\tau)\rangle d\tau$, can be simplified for such model at the anti-Stokes peak, $\omega_\mathrm{aS}=\omega_L+\omega_v$, as follows (see derivation in SM~\cite{supp3})
\begin{align}
	\label{eqn:coefficients}
S^\mathrm{theo} _\mathrm{R,aS}(\omega_L+\omega_v)=&|C^\mathrm{res}_{\mathrm{aS}}+C^\mathrm{off}_{\mathrm{aS}}|^2\times\\ \nonumber
&|C^v_{\mathrm{aS}}\left(\langle v^\dagger v\rangle,\omega_L+\omega_v\right)|^2,
\end{align}
	where $C^\mathrm{off}_{\mathrm{aS}}=	i g_{\mathrm{om}} {\alpha_s}/ [ i(\Delta^\prime_c-\omega_v)+\frac{\kappa}{2} ]$ corresponds to the off-resonant Raman amplitude and   $C^\mathrm{res}_{\mathrm{aS}}=	 -g g_{0,r} \mean{\xi_r}/[( i(\Delta_r-\omega_v)+\frac{\Gamma_\mathrm{eff}}{2}) ( i(\Delta_c^\prime-\omega_v)+\frac{\kappa}{2} )]$ to the on-resonance one, proportional to their respective Raman polarizabilities. The Purcell-enhanced decay rate of the emitter is denoted by $\Gamma_\mathrm{eff}=\frac{g²\kappa}{(\kappa/2)²+(\Delta_r-\Delta_c^\prime)² }$, that modifies the emitter induced coherent amplitude $\langle\hat\xi_r\rangle\approx-ig_r{\alpha_s}/[i\Delta_r+\Gamma_\mathrm{eff}/2]$ and ${\alpha_s}=-\Omega/(\Delta_c-i(k/2))$ denotes the driven amplitude for the bare cavity~\cite{Neuman2019}.
The sum of these coefficients allows us to distinguish between off and near-resonant coherent contributions to the anti-Stokes Raman peak, together with the spectral contribution of the vibron correlation $|C^v_{{\mathrm{aS}}}\left(\langle v^\dagger v\rangle,\omega\right)|^2=\gamma_v \mean{v^\dagger v}/[(\omega_L+\omega_v-\omega)^2+(\frac{\gamma_v}{2})^2]$, which accounts for both. Although we have focused on the anti-Stokes lines, we also point out that similar expressions and coherent effects can be observed for the Stokes emission (see SM~\cite{supp3}).
\begin{figure}
	\begin{center}
		\includegraphics[width=8.6cm]{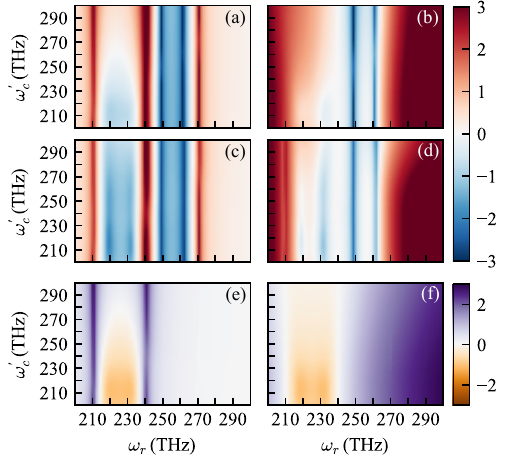}
	\end{center}
     \vspace{-0.2cm}
	\caption{(a)-(b) Semi-analytical anti-Stokes spectral ratio of the total model $S_\mathrm{R}(\omega_L+\omega_v)$, using formula~(\ref{eqn:coefficients}), compared to the standard optomechanical model, $\mathrm{log}\left[S^\mathrm{theo}_\mathrm{R,aS}/S^\mathrm{om,theo}_\mathrm{R,aS}\right]$  (panel a), and compared to the bare resonant one, $\mathrm{log}\left[S^\mathrm{theo}_\mathrm{R,aS}/S^\mathrm{res,theo}_\mathrm{R,aS}\right]$ (panel b). (c)-(d) Analogous full-computational anti-Stokes Raman ratios, $\mathrm{log}\left[S_\mathrm{aS}/S^\mathrm{om}_\mathrm{aS}\right]$ (panel c), and  $\mathrm{log}\left[S_\mathrm{aS}/S^\mathrm{res}_\mathrm{aS}\right]$ (panel d), respectively. (e) Vibron population ratio comparing the total model with the optomechanical one, $\mathrm{log}\left[\langle v^\dagger v\rangle/\langle v^\dagger v\rangle_\mathrm{om}\right]$, and (f) compared to the bare resonant model, $\mathrm{log}\left[\langle v^\dagger v\rangle/\langle v^\dagger v\rangle_\mathrm{res}\right]$. The rest of the parameters coincide with Fig.~\ref{CooperativeRamanNearandOff}. }
	\label{CooperativeMaps}
	\vspace{-0.2cm}
\end{figure}

In Fig.~(\ref{CooperativeMaps})(a)-(b)  we show the analytical ratios of the anti-Stokes peaks between the total model $H_{\mathrm{om+res}}$ and the standard optomechanical model $H_{\mathrm{om}}$ (panel a), and the bare models for resonance SERS $H_{\mathrm{res}}$ (panel b), as a function of the cavity and near-resonant transition frequencies. Analogously, panels (c) and (d) display the similar ratios for the full-computational results, which show a remarkable agreement with the simplified analytical model.
The spectral regions with coincident enhancements or suppressions on the left and right panels allow us identifying coherent Raman interference between off-resonant and near-resonant transitions, which are confirmed by the analytical model.  These interferences arise from the comparable values of the near-resonant and the off-resonant Raman polarizabilities, which occur for vibrations with significant Raman activities at frequency regions where the ZPL is close to the Stokes, the laser or the anti-Stokes resonance, respectively. We notice that such phenomena is further enhanced optomechanically by the vibrational population assisted by the near-resonance and off-resonance electronic transitions (see panels (e)-(f) and the vibrational contribution in Eq.~(\ref{eqn:coefficients})). The effect of room temperature baths results in a predominant thermal population over the optomechanically induced one, which can mostly suppress such vibrational Raman enhancements (see SM~\cite{supp3}). Furthermore, for molecular vibrations with smaller Raman activities, we point out that the near-resonant Raman polarizability dominates over the off-resonant one and the cavity spectra between the bare resonance SERS and the combined model conveys to the same values in accordance with previous results \cite{Neuman2019}. 

\begin{figure}
	\begin{center}
		\includegraphics[width=8.6cm]{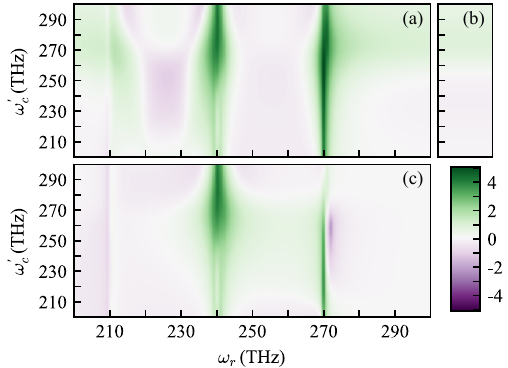}
	\end{center}
\vspace{-0.2cm}
	\caption{Map of the Cauchy-Schwarz  inequality parameter $R_\Gamma(\omega_\mathrm{S},\omega_\mathrm{aS})$ in logarithmic scale, for the optomechanical Hamiltonian $H_\mathrm{om}$~\cite{Roelli2016} (b), the resonant SERS Hamiltonian $H_\mathrm{res}$~\cite{Neuman2019} (c), and the total Hamiltonian $H_{\mathrm{om+ res}}$ containing both contributions (a). The filter linewidth is $\Gamma=\kappa_v$. The parameters coincide with those in Fig.~\ref{CooperativeMaps}.}
	\label{Quantumcorrelations}
	\vspace{-0.6cm}
\end{figure}

The quantum optomechanical description of SERS triggers the interest of exploring evidences of nonclassical features~\cite{SanchezMunoz2014} arising from the cooperative Raman processes described above. Previous  works have proposed the generation of quantum correlated pairs of Stokes and anti-Stokes photons~\cite{Klyshko1977}, that were confirmed  in experiments~\cite{Kasperczyk2016,Anderson2018}, and can be severely modified by strong optomechanical couplings in SERS~\cite{Schmidt2016,Schmidt2021}.
Exploiting the sensor method~\cite{SanchezMunoz2014}, we compute in Fig.\ref{Quantumcorrelations} the Cauchy-Schwarz inequality (CSI) $R=(g^{(2)}_\Gamma(\omega_\mathrm{S},\omega_\mathrm{aS}))^2/(g^{(2)}_\Gamma(\omega_\mathrm{S})g^{(2)}_\Gamma(\omega_\mathrm{aS}))$ expressed via the frequency-filtered second-order photon correlations between the Stokes and anti-Stokes, $g^{(2)}_\Gamma(\omega_\mathrm{S},\omega_\mathrm{aS})$,  as a function of  the cavity and ZPL frequencies. For cooperative processes with the ZPL  near the anti-Stokes peak (a), we identify large nonclassical  violations of the CSI~\cite{SanchezMunoz2014} ($R\approx10^{5}\gg1$). Such quantum violation is increased up to five orders of magnitude compared to the previous standard models (b, c), which originates from the cooperative optomechanical interaction between the near and off-resonant SERS that enhances the generation of Stokes-anti-Stokes photon pairs (see $g^{(2)}$ maps in SM~\cite{supp3}).

In conclusion, we have proposed a molecular optomechanical model from first microscopic principles that allows both, mapping to the standard molecular optomechanical Hamiltonian in SERS for off-resonant electronic levels, and accounting non-adiabatic effects of near resonant electronic transitions at the same foot level. We have shown that the inclusion of both transitions kinds enables unique interference mechanisms that modifies the predictions of off-resonance and resonance SERS independent models.  Overall these results point out to the subtleties of measuring the optomechanical coupling constant from anti-Stokes-Stokes thermometry at the moderate drivings used in SERS~\cite{Benz2016} as it is standard in cavity optomechanics~\cite{Aspelmeyer2014}. Such cooperative modifications of the Raman peaks call for experimental evidence by combining standard spectral SERS techniques~\cite{Benz2016,Chen2021,Chen2021a}  and  quantum correlation measurements~\cite{Kasperczyk2016,Anderson2018} with state-of-the-art ZPL tuning techniques for quantum emitters~\cite{Toninelli2021}. Whereas in this work we considered strong Raman activities and enhanced optomechanical cooperativities due to the presence of a small volume plasmonic cavity, we notice that such coherent effects could be also accessed similarly in nearly lossless dielectric nanoparticles~\cite{caldarola2015,albella2013,kuznetsov2016} and weaker Raman scenarios or free-space at stronger laser intensities or different detuning parameters.These results call for interfacing more complete evaluations of the molecular spectral properties~\cite{Guthmuller2016,Kumar1992} in combination with numerically exact methods~\cite{delPino2018,Fowler-Wright2022,Cygorek2022}, to explore the full non-Markovian dynamics of the molecular optomechanical Hamiltonian, including vibrational nonlinear phenomena at larger displacements that can induce modified SERS parametric instabilities~\cite{Roelli2016} and chemical reactions~\cite{Zrimsek2016}. Another future perspective is to research the cooperative dynamics here shown in the presence of collective molecular interactions to  search new fundamental limits and cooperative mechanisms in optomechanics \cite{Aspelmeyer2014}.

\begin{acknowledgements}{All the simulations in this work were obtained with the Quantum Toolbox in
		Python (QuTiP) \cite{Johansson2012}. We acknowledge financial support from the  Ramon y Cajal program (RYC2020-029730-I) and the 
		fellowship (LCF/BQ/PI20/11760018) from ''La Caixa''
		Foundation (ID 100010434) and from the European
		Union's Horizon 2020 research and innovation program
		under the Marie Sklodowska-Curie Grant No. 847648. We also acknowledge financial
		support from the MCINN projects PID2021-126964OB-I00 (QENIGMA) and TED2021-130552B-C21 (ADIQUNANO), and the CAM project  Sin\'ergico CAM 2020 Y2020/TCS-
		6545 (NanoQuCo-CM). We also acknowledge Caspar Groiseau and Johannes Feist for fruitful discussions.}
\end{acknowledgements}
\bibliographystyle{apsrev4-2}
\bibliography{Electronic_dynamics_SERS}

\begin{thebibliography}{51}%
\makeatletter
\providecommand \@ifxundefined [1]{%
 \@ifx{#1\undefined}
}%
\providecommand \@ifnum [1]{%
 \ifnum #1\expandafter \@firstoftwo
 \else \expandafter \@secondoftwo
 \fi
}%
\providecommand \@ifx [1]{%
 \ifx #1\expandafter \@firstoftwo
 \else \expandafter \@secondoftwo
 \fi
}%
\providecommand \natexlab [1]{#1}%
\providecommand \enquote  [1]{``#1''}%
\providecommand \bibnamefont  [1]{#1}%
\providecommand \bibfnamefont [1]{#1}%
\providecommand \citenamefont [1]{#1}%
\providecommand \href@noop [0]{\@secondoftwo}%
\providecommand \href [0]{\begingroup \@sanitize@url \@href}%
\providecommand \@href[1]{\@@startlink{#1}\@@href}%
\providecommand \@@href[1]{\endgroup#1\@@endlink}%
\providecommand \@sanitize@url [0]{\catcode `\\12\catcode `\$12\catcode
  `\&12\catcode `\#12\catcode `\^12\catcode `\_12\catcode `\%12\relax}%
\providecommand \@@startlink[1]{}%
\providecommand \@@endlink[0]{}%
\providecommand \url  [0]{\begingroup\@sanitize@url \@url }%
\providecommand \@url [1]{\endgroup\@href {#1}{\urlprefix }}%
\providecommand \urlprefix  [0]{URL }%
\providecommand \Eprint [0]{\href }%
\providecommand \doibase [0]{https://doi.org/}%
\providecommand \selectlanguage [0]{\@gobble}%
\providecommand \bibinfo  [0]{\@secondoftwo}%
\providecommand \bibfield  [0]{\@secondoftwo}%
\providecommand \translation [1]{[#1]}%
\providecommand \BibitemOpen [0]{}%
\providecommand \bibitemStop [0]{}%
\providecommand \bibitemNoStop [0]{.\EOS\space}%
\providecommand \EOS [0]{\spacefactor3000\relax}%
\providecommand \BibitemShut  [1]{\csname bibitem#1\endcsname}%
\let\auto@bib@innerbib\@empty
\bibitem [{\citenamefont {Fleischmann}\ \emph {et~al.}(1974)\citenamefont
  {Fleischmann}, \citenamefont {Hendra},\ and\ \citenamefont
  {McQuillan}}]{Fleischmann1974}%
  \BibitemOpen
  \bibfield  {author} {\bibinfo {author} {\bibfnamefont {M.}~\bibnamefont
  {Fleischmann}}, \bibinfo {author} {\bibfnamefont {P.~J.}\ \bibnamefont
  {Hendra}},\ and\ \bibinfo {author} {\bibfnamefont {A.~J.}\ \bibnamefont
  {McQuillan}},\ }\href@noop {} {\bibfield  {journal} {\bibinfo  {journal}
  {Chem. Phys. Lett.}\ }\textbf {\bibinfo {volume} {26}},\ \bibinfo {pages}
  {163} (\bibinfo {year} {1974})}\BibitemShut {NoStop}%
\bibitem [{\citenamefont {Jeanmaire}\ and\ \citenamefont
  {Van~Duyne}(1977)}]{Jeanmaire1977}%
  \BibitemOpen
  \bibfield  {author} {\bibinfo {author} {\bibfnamefont {D.~L.}\ \bibnamefont
  {Jeanmaire}}\ and\ \bibinfo {author} {\bibfnamefont {R.~P.}\ \bibnamefont
  {Van~Duyne}},\ }\href {https://doi.org/10.1016/S0022-0728(77)80224-6}
  {\bibfield  {journal} {\bibinfo  {journal} {Journal of Electroanalytical
  Chemistry and Interfacial Electrochemistry}\ }\textbf {\bibinfo {volume}
  {84}},\ \bibinfo {pages} {1} (\bibinfo {year} {1977})}\BibitemShut {NoStop}%
\bibitem [{\citenamefont {Albrecht}\ and\ \citenamefont
  {Creighton}(1977)}]{Albrecht1977}%
  \BibitemOpen
  \bibfield  {author} {\bibinfo {author} {\bibfnamefont {M.~G.}\ \bibnamefont
  {Albrecht}}\ and\ \bibinfo {author} {\bibfnamefont {J.~A.}\ \bibnamefont
  {Creighton}},\ }\href@noop {} {\bibfield  {journal} {\bibinfo  {journal} {J.
  Am. Chem. Soc.}\ }\textbf {\bibinfo {volume} {99}},\ \bibinfo {pages} {5215}
  (\bibinfo {year} {1977})}\BibitemShut {NoStop}%
\bibitem [{\citenamefont {Kneipp}\ \emph {et~al.}(1997)\citenamefont {Kneipp},
  \citenamefont {Wang}, \citenamefont {Kneipp}, \citenamefont {Perelman},
  \citenamefont {Itzkan}, \citenamefont {Dasari},\ and\ \citenamefont
  {Feld}}]{Kneipp1997}%
  \BibitemOpen
  \bibfield  {author} {\bibinfo {author} {\bibfnamefont {K.}~\bibnamefont
  {Kneipp}}, \bibinfo {author} {\bibfnamefont {Y.}~\bibnamefont {Wang}},
  \bibinfo {author} {\bibfnamefont {H.}~\bibnamefont {Kneipp}}, \bibinfo
  {author} {\bibfnamefont {L.~T.}\ \bibnamefont {Perelman}}, \bibinfo {author}
  {\bibfnamefont {I.}~\bibnamefont {Itzkan}}, \bibinfo {author} {\bibfnamefont
  {R.~R.}\ \bibnamefont {Dasari}},\ and\ \bibinfo {author} {\bibfnamefont
  {M.~S.}\ \bibnamefont {Feld}},\ }\href@noop {} {\bibfield  {journal}
  {\bibinfo  {journal} {Phys. Rev. Lett.}\ }\textbf {\bibinfo {volume} {78}},\
  \bibinfo {pages} {1667} (\bibinfo {year} {1997})}\BibitemShut {NoStop}%
\bibitem [{\citenamefont {Nie}\ and\ \citenamefont {Emory}(1997)}]{Nie1997}%
  \BibitemOpen
  \bibfield  {author} {\bibinfo {author} {\bibfnamefont {S.}~\bibnamefont
  {Nie}}\ and\ \bibinfo {author} {\bibfnamefont {S.}~\bibnamefont {Emory}},\
  }\href@noop {} {\bibfield  {journal} {\bibinfo  {journal} {Science}\ }\textbf
  {\bibinfo {volume} {275}},\ \bibinfo {pages} {1102} (\bibinfo {year}
  {1997})}\BibitemShut {NoStop}%
\bibitem [{\citenamefont {Zrimsek}\ \emph {et~al.}(2016)\citenamefont
  {Zrimsek}, \citenamefont {Chiang}, \citenamefont {Mattei}, \citenamefont
  {Zaleski}, \citenamefont {McAnally}, \citenamefont {Chapman}, \citenamefont
  {Henry}, \citenamefont {Schatz},\ and\ \citenamefont
  {Van~Duyne}}]{Zrimsek2016}%
  \BibitemOpen
  \bibfield  {author} {\bibinfo {author} {\bibfnamefont {A.~B.}\ \bibnamefont
  {Zrimsek}}, \bibinfo {author} {\bibfnamefont {N.}~\bibnamefont {Chiang}},
  \bibinfo {author} {\bibfnamefont {M.}~\bibnamefont {Mattei}}, \bibinfo
  {author} {\bibfnamefont {S.}~\bibnamefont {Zaleski}}, \bibinfo {author}
  {\bibfnamefont {M.~O.}\ \bibnamefont {McAnally}}, \bibinfo {author}
  {\bibfnamefont {C.~T.}\ \bibnamefont {Chapman}}, \bibinfo {author}
  {\bibfnamefont {A.-I.}\ \bibnamefont {Henry}}, \bibinfo {author}
  {\bibfnamefont {G.~C.}\ \bibnamefont {Schatz}},\ and\ \bibinfo {author}
  {\bibfnamefont {R.~P.}\ \bibnamefont {Van~Duyne}},\ }\href@noop {} {\bibfield
   {journal} {\bibinfo  {journal} {Chemical Reviews}\ }\textbf {\bibinfo
  {volume} {117}},\ \bibinfo {pages} {7583} (\bibinfo {year}
  {2016})}\BibitemShut {NoStop}%
\bibitem [{\citenamefont {Moskovits}(2013)}]{Moskovits2013}%
  \BibitemOpen
  \bibfield  {author} {\bibinfo {author} {\bibfnamefont {M.}~\bibnamefont
  {Moskovits}},\ }\href {https://doi.org/10.1039/C2CP44030J} {\bibfield
  {journal} {\bibinfo  {journal} {Physical Chemistry Chemical Physics}\
  }\textbf {\bibinfo {volume} {15}},\ \bibinfo {pages} {5301} (\bibinfo {year}
  {2013})}\BibitemShut {NoStop}%
\bibitem [{\citenamefont {Roelli}\ \emph {et~al.}(2016)\citenamefont {Roelli},
  \citenamefont {Galland}, \citenamefont {Piro},\ and\ \citenamefont
  {Kippenberg}}]{Roelli2016}%
  \BibitemOpen
  \bibfield  {author} {\bibinfo {author} {\bibfnamefont {P.}~\bibnamefont
  {Roelli}}, \bibinfo {author} {\bibfnamefont {C.}~\bibnamefont {Galland}},
  \bibinfo {author} {\bibfnamefont {N.}~\bibnamefont {Piro}},\ and\ \bibinfo
  {author} {\bibfnamefont {T.~J.}\ \bibnamefont {Kippenberg}},\ }\href
  {https://doi.org/10.1038/nnano.2015.264} {\bibfield  {journal} {\bibinfo
  {journal} {Nature Nanotechnology}\ }\textbf {\bibinfo {volume} {11}},\
  \bibinfo {pages} {164} (\bibinfo {year} {2016})}\BibitemShut {NoStop}%
\bibitem [{\citenamefont {Aspelmeyer}\ \emph {et~al.}(2014)\citenamefont
  {Aspelmeyer}, \citenamefont {Kippenberg},\ and\ \citenamefont
  {Marquardt}}]{Aspelmeyer2014}%
  \BibitemOpen
  \bibfield  {author} {\bibinfo {author} {\bibfnamefont {M.}~\bibnamefont
  {Aspelmeyer}}, \bibinfo {author} {\bibfnamefont {T.~J.}\ \bibnamefont
  {Kippenberg}},\ and\ \bibinfo {author} {\bibfnamefont {F.}~\bibnamefont
  {Marquardt}},\ }\href@noop {} {\bibfield  {journal} {\bibinfo  {journal}
  {Rev. Mod. Phys.}\ }\textbf {\bibinfo {volume} {86}},\ \bibinfo {pages}
  {1391} (\bibinfo {year} {2014})}\BibitemShut {NoStop}%
\bibitem [{\citenamefont {Esteban}\ \emph {et~al.}(2022)\citenamefont
  {Esteban}, \citenamefont {Baumberg},\ and\ \citenamefont
  {Aizpurua}}]{Esteban2022}%
  \BibitemOpen
  \bibfield  {author} {\bibinfo {author} {\bibfnamefont {R.}~\bibnamefont
  {Esteban}}, \bibinfo {author} {\bibfnamefont {J.~J.}\ \bibnamefont
  {Baumberg}},\ and\ \bibinfo {author} {\bibfnamefont {J.}~\bibnamefont
  {Aizpurua}},\ }\href {https://doi.org/10.1021/acs.accounts.1c00759}
  {\bibfield  {journal} {\bibinfo  {journal} {Acc. Chem. Res.}\ }\textbf
  {\bibinfo {volume} {55}},\ \bibinfo {pages} {1889} (\bibinfo {year}
  {2022})}\BibitemShut {NoStop}%
\bibitem [{\citenamefont {Schmidt}\ \emph {et~al.}(2016)\citenamefont
  {Schmidt}, \citenamefont {Esteban}, \citenamefont {{Gonz{\'a}lez-Tudela}},
  \citenamefont {Giedke},\ and\ \citenamefont {Aizpurua}}]{Schmidt2016}%
  \BibitemOpen
  \bibfield  {author} {\bibinfo {author} {\bibfnamefont {M.~K.}\ \bibnamefont
  {Schmidt}}, \bibinfo {author} {\bibfnamefont {R.}~\bibnamefont {Esteban}},
  \bibinfo {author} {\bibfnamefont {A.}~\bibnamefont {{Gonz{\'a}lez-Tudela}}},
  \bibinfo {author} {\bibfnamefont {G.}~\bibnamefont {Giedke}},\ and\ \bibinfo
  {author} {\bibfnamefont {J.}~\bibnamefont {Aizpurua}},\ }\href
  {https://doi.org/10.1021/acsnano.6b02484} {\bibfield  {journal} {\bibinfo
  {journal} {ACS Nano}\ }\textbf {\bibinfo {volume} {10}},\ \bibinfo {pages}
  {6291} (\bibinfo {year} {2016})}\BibitemShut {NoStop}%
\bibitem [{\citenamefont {Roelli}\ \emph {et~al.}(2020)\citenamefont {Roelli},
  \citenamefont {{Martin-Cano}}, \citenamefont {Kippenberg},\ and\
  \citenamefont {Galland}}]{Roelli2020}%
  \BibitemOpen
  \bibfield  {author} {\bibinfo {author} {\bibfnamefont {P.}~\bibnamefont
  {Roelli}}, \bibinfo {author} {\bibfnamefont {D.}~\bibnamefont
  {{Martin-Cano}}}, \bibinfo {author} {\bibfnamefont {T.~J.}\ \bibnamefont
  {Kippenberg}},\ and\ \bibinfo {author} {\bibfnamefont {C.}~\bibnamefont
  {Galland}},\ }\href {https://doi.org/10.1103/PhysRevX.10.031057} {\bibfield
  {journal} {\bibinfo  {journal} {Phys. Rev. X}\ }\textbf {\bibinfo {volume}
  {10}},\ \bibinfo {pages} {031057} (\bibinfo {year} {2020})}\BibitemShut
  {NoStop}%
\bibitem [{\citenamefont {Neuman}\ \emph {et~al.}(2019)\citenamefont {Neuman},
  \citenamefont {Esteban}, \citenamefont {Giedke}, \citenamefont {Schmidt},\
  and\ \citenamefont {Aizpurua}}]{Neuman2019}%
  \BibitemOpen
  \bibfield  {author} {\bibinfo {author} {\bibfnamefont {T.}~\bibnamefont
  {Neuman}}, \bibinfo {author} {\bibfnamefont {R.}~\bibnamefont {Esteban}},
  \bibinfo {author} {\bibfnamefont {G.}~\bibnamefont {Giedke}}, \bibinfo
  {author} {\bibfnamefont {M.~K.}\ \bibnamefont {Schmidt}},\ and\ \bibinfo
  {author} {\bibfnamefont {J.}~\bibnamefont {Aizpurua}},\ }\href
  {https://doi.org/10.1103/PhysRevA.100.043422} {\bibfield  {journal} {\bibinfo
   {journal} {Phys. Rev. A}\ }\textbf {\bibinfo {volume} {100}},\ \bibinfo
  {pages} {043422} (\bibinfo {year} {2019})}\BibitemShut {NoStop}%
\bibitem [{\citenamefont {Neuman}\ \emph {et~al.}(2020)\citenamefont {Neuman},
  \citenamefont {Aizpurua},\ and\ \citenamefont {Esteban}}]{Neuman2020}%
  \BibitemOpen
  \bibfield  {author} {\bibinfo {author} {\bibfnamefont {T.}~\bibnamefont
  {Neuman}}, \bibinfo {author} {\bibfnamefont {J.}~\bibnamefont {Aizpurua}},\
  and\ \bibinfo {author} {\bibfnamefont {R.}~\bibnamefont {Esteban}},\ }\href
  {https://doi.org/10.1515/nanoph-2019-0336} {\bibfield  {journal} {\bibinfo
  {journal} {Nanophotonics}\ }\textbf {\bibinfo {volume} {9}},\ \bibinfo
  {pages} {295} (\bibinfo {year} {2020})}\BibitemShut {NoStop}%
\bibitem [{\citenamefont {Dezfouli}\ \emph {et~al.}(2019)\citenamefont
  {Dezfouli}, \citenamefont {Gordon},\ and\ \citenamefont
  {Hughes}}]{Dezfouli2019}%
  \BibitemOpen
  \bibfield  {author} {\bibinfo {author} {\bibfnamefont {M.~K.}\ \bibnamefont
  {Dezfouli}}, \bibinfo {author} {\bibfnamefont {R.}~\bibnamefont {Gordon}},\
  and\ \bibinfo {author} {\bibfnamefont {S.}~\bibnamefont {Hughes}},\ }\href
  {https://doi.org/10.1021/acsphotonics.8b01091} {\bibfield  {journal}
  {\bibinfo  {journal} {ACS Photonics}\ }\textbf {\bibinfo {volume} {6}},\
  \bibinfo {pages} {1400} (\bibinfo {year} {2019})}\BibitemShut {NoStop}%
\bibitem [{\citenamefont {Gurlek}\ \emph {et~al.}(2021)\citenamefont {Gurlek},
  \citenamefont {Sandoghdar},\ and\ \citenamefont
  {{Martin-Cano}}}]{Gurlek2021}%
  \BibitemOpen
  \bibfield  {author} {\bibinfo {author} {\bibfnamefont {B.}~\bibnamefont
  {Gurlek}}, \bibinfo {author} {\bibfnamefont {V.}~\bibnamefont {Sandoghdar}},\
  and\ \bibinfo {author} {\bibfnamefont {D.}~\bibnamefont {{Martin-Cano}}},\
  }\href {https://doi.org/10.1103/PhysRevLett.127.123603} {\bibfield  {journal}
  {\bibinfo  {journal} {Phys. Rev. Lett.}\ }\textbf {\bibinfo {volume} {127}},\
  \bibinfo {pages} {123603} (\bibinfo {year} {2021})}\BibitemShut {NoStop}%
\bibitem [{\citenamefont {Benz}\ \emph {et~al.}(2016)\citenamefont {Benz},
  \citenamefont {Schmidt}, \citenamefont {Dreismann}, \citenamefont
  {Chikkaraddy}, \citenamefont {Zhang}, \citenamefont {Demetriadou},
  \citenamefont {Carnegie}, \citenamefont {Ohadi}, \citenamefont {{de Nijs}},
  \citenamefont {Esteban}, \citenamefont {Aizpurua},\ and\ \citenamefont
  {Baumberg}}]{Benz2016}%
  \BibitemOpen
  \bibfield  {author} {\bibinfo {author} {\bibfnamefont {F.}~\bibnamefont
  {Benz}}, \bibinfo {author} {\bibfnamefont {M.~K.}\ \bibnamefont {Schmidt}},
  \bibinfo {author} {\bibfnamefont {A.}~\bibnamefont {Dreismann}}, \bibinfo
  {author} {\bibfnamefont {R.}~\bibnamefont {Chikkaraddy}}, \bibinfo {author}
  {\bibfnamefont {Y.}~\bibnamefont {Zhang}}, \bibinfo {author} {\bibfnamefont
  {A.}~\bibnamefont {Demetriadou}}, \bibinfo {author} {\bibfnamefont
  {C.}~\bibnamefont {Carnegie}}, \bibinfo {author} {\bibfnamefont
  {H.}~\bibnamefont {Ohadi}}, \bibinfo {author} {\bibfnamefont
  {B.}~\bibnamefont {{de Nijs}}}, \bibinfo {author} {\bibfnamefont
  {R.}~\bibnamefont {Esteban}}, \bibinfo {author} {\bibfnamefont
  {J.}~\bibnamefont {Aizpurua}},\ and\ \bibinfo {author} {\bibfnamefont
  {J.~J.}\ \bibnamefont {Baumberg}},\ }\href
  {https://doi.org/10.1126/science.aah5243} {\bibfield  {journal} {\bibinfo
  {journal} {Science}\ }\textbf {\bibinfo {volume} {354}},\ \bibinfo {pages}
  {726} (\bibinfo {year} {2016})}\BibitemShut {NoStop}%
\bibitem [{\citenamefont {Lombardi}\ \emph {et~al.}(2018)\citenamefont
  {Lombardi}, \citenamefont {Schmidt}, \citenamefont {Weller}, \citenamefont
  {Deacon}, \citenamefont {Benz}, \citenamefont {{de Nijs}}, \citenamefont
  {Aizpurua},\ and\ \citenamefont {Baumberg}}]{Lombardi2018}%
  \BibitemOpen
  \bibfield  {author} {\bibinfo {author} {\bibfnamefont {A.}~\bibnamefont
  {Lombardi}}, \bibinfo {author} {\bibfnamefont {M.~K.}\ \bibnamefont
  {Schmidt}}, \bibinfo {author} {\bibfnamefont {L.}~\bibnamefont {Weller}},
  \bibinfo {author} {\bibfnamefont {W.~M.}\ \bibnamefont {Deacon}}, \bibinfo
  {author} {\bibfnamefont {F.}~\bibnamefont {Benz}}, \bibinfo {author}
  {\bibfnamefont {B.}~\bibnamefont {{de Nijs}}}, \bibinfo {author}
  {\bibfnamefont {J.}~\bibnamefont {Aizpurua}},\ and\ \bibinfo {author}
  {\bibfnamefont {J.~J.}\ \bibnamefont {Baumberg}},\ }\href@noop {} {\bibfield
  {journal} {\bibinfo  {journal} {Phys. Rev. X}\ }\textbf {\bibinfo {volume}
  {8}},\ \bibinfo {pages} {011016} (\bibinfo {year} {2018})}\BibitemShut
  {NoStop}%
\bibitem [{\citenamefont {Chen}\ \emph
  {et~al.}(2021{\natexlab{a}})\citenamefont {Chen}, \citenamefont {Roelli},
  \citenamefont {Ahmed}, \citenamefont {Verlekar}, \citenamefont {Hu},
  \citenamefont {Banjac}, \citenamefont {Lingenfelder}, \citenamefont
  {Kippenberg}, \citenamefont {Tagliabue},\ and\ \citenamefont
  {Galland}}]{Chen2021a}%
  \BibitemOpen
  \bibfield  {author} {\bibinfo {author} {\bibfnamefont {W.}~\bibnamefont
  {Chen}}, \bibinfo {author} {\bibfnamefont {P.}~\bibnamefont {Roelli}},
  \bibinfo {author} {\bibfnamefont {A.}~\bibnamefont {Ahmed}}, \bibinfo
  {author} {\bibfnamefont {S.}~\bibnamefont {Verlekar}}, \bibinfo {author}
  {\bibfnamefont {H.}~\bibnamefont {Hu}}, \bibinfo {author} {\bibfnamefont
  {K.}~\bibnamefont {Banjac}}, \bibinfo {author} {\bibfnamefont
  {M.}~\bibnamefont {Lingenfelder}}, \bibinfo {author} {\bibfnamefont {T.~J.}\
  \bibnamefont {Kippenberg}}, \bibinfo {author} {\bibfnamefont
  {G.}~\bibnamefont {Tagliabue}},\ and\ \bibinfo {author} {\bibfnamefont
  {C.}~\bibnamefont {Galland}},\ }\href
  {https://doi.org/10.1038/s41467-021-22679-y} {\bibfield  {journal} {\bibinfo
  {journal} {Nat Commun}\ }\textbf {\bibinfo {volume} {12}},\ \bibinfo {pages}
  {2731} (\bibinfo {year} {2021}{\natexlab{a}})}\BibitemShut {NoStop}%
\bibitem [{\citenamefont {Xomalis}\ \emph {et~al.}(2021)\citenamefont
  {Xomalis}, \citenamefont {Zheng}, \citenamefont {Chikkaraddy}, \citenamefont
  {{Koczor-Benda}}, \citenamefont {Miele}, \citenamefont {Rosta}, \citenamefont
  {Vandenbosch}, \citenamefont {Mart{\'i}nez},\ and\ \citenamefont
  {Baumberg}}]{Xomalis2021}%
  \BibitemOpen
  \bibfield  {author} {\bibinfo {author} {\bibfnamefont {A.}~\bibnamefont
  {Xomalis}}, \bibinfo {author} {\bibfnamefont {X.}~\bibnamefont {Zheng}},
  \bibinfo {author} {\bibfnamefont {R.}~\bibnamefont {Chikkaraddy}}, \bibinfo
  {author} {\bibfnamefont {Z.}~\bibnamefont {{Koczor-Benda}}}, \bibinfo
  {author} {\bibfnamefont {E.}~\bibnamefont {Miele}}, \bibinfo {author}
  {\bibfnamefont {E.}~\bibnamefont {Rosta}}, \bibinfo {author} {\bibfnamefont
  {G.~A.~E.}\ \bibnamefont {Vandenbosch}}, \bibinfo {author} {\bibfnamefont
  {A.}~\bibnamefont {Mart{\'i}nez}},\ and\ \bibinfo {author} {\bibfnamefont
  {J.~J.}\ \bibnamefont {Baumberg}},\ }\href
  {https://doi.org/10.1126/science.abk2593} {\bibfield  {journal} {\bibinfo
  {journal} {Science}\ }\textbf {\bibinfo {volume} {374}},\ \bibinfo {pages}
  {1268} (\bibinfo {year} {2021})}\BibitemShut {NoStop}%
\bibitem [{\citenamefont {Chen}\ \emph
  {et~al.}(2021{\natexlab{b}})\citenamefont {Chen}, \citenamefont {Roelli},
  \citenamefont {Hu}, \citenamefont {Verlekar}, \citenamefont {Amirtharaj},
  \citenamefont {Barreda}, \citenamefont {Kippenberg}, \citenamefont
  {Kovylina}, \citenamefont {Verhagen}, \citenamefont {Mart{\'i}nez},\ and\
  \citenamefont {Galland}}]{Chen2021}%
  \BibitemOpen
  \bibfield  {author} {\bibinfo {author} {\bibfnamefont {W.}~\bibnamefont
  {Chen}}, \bibinfo {author} {\bibfnamefont {P.}~\bibnamefont {Roelli}},
  \bibinfo {author} {\bibfnamefont {H.}~\bibnamefont {Hu}}, \bibinfo {author}
  {\bibfnamefont {S.}~\bibnamefont {Verlekar}}, \bibinfo {author}
  {\bibfnamefont {S.~P.}\ \bibnamefont {Amirtharaj}}, \bibinfo {author}
  {\bibfnamefont {A.~I.}\ \bibnamefont {Barreda}}, \bibinfo {author}
  {\bibfnamefont {T.~J.}\ \bibnamefont {Kippenberg}}, \bibinfo {author}
  {\bibfnamefont {M.}~\bibnamefont {Kovylina}}, \bibinfo {author}
  {\bibfnamefont {E.}~\bibnamefont {Verhagen}}, \bibinfo {author}
  {\bibfnamefont {A.}~\bibnamefont {Mart{\'i}nez}},\ and\ \bibinfo {author}
  {\bibfnamefont {C.}~\bibnamefont {Galland}},\ }\href
  {https://doi.org/10.1126/science.abk3106} {\bibfield  {journal} {\bibinfo
  {journal} {Science}\ }\textbf {\bibinfo {volume} {374}},\ \bibinfo {pages}
  {1264} (\bibinfo {year} {2021}{\natexlab{b}})}\BibitemShut {NoStop}%
\bibitem [{\citenamefont {Weiss}\ \emph {et~al.}(2013)\citenamefont {Weiss},
  \citenamefont {Bruder},\ and\ \citenamefont {Nunnenkamp}}]{Weiss2013a}%
  \BibitemOpen
  \bibfield  {author} {\bibinfo {author} {\bibfnamefont {T.}~\bibnamefont
  {Weiss}}, \bibinfo {author} {\bibfnamefont {C.}~\bibnamefont {Bruder}},\ and\
  \bibinfo {author} {\bibfnamefont {A.}~\bibnamefont {Nunnenkamp}},\ }\href
  {https://doi.org/10.1088/1367-2630/15/4/045017} {\bibfield  {journal}
  {\bibinfo  {journal} {New J. Phys.}\ }\textbf {\bibinfo {volume} {15}},\
  \bibinfo {pages} {045017} (\bibinfo {year} {2013})}\BibitemShut {NoStop}%
\bibitem [{\citenamefont {{Wilson-Rae}}\ \emph {et~al.}(2008)\citenamefont
  {{Wilson-Rae}}, \citenamefont {Nooshi}, \citenamefont {Dobrindt},
  \citenamefont {Kippenberg},\ and\ \citenamefont {Zwerger}}]{Wilson-Rae2008}%
  \BibitemOpen
  \bibfield  {author} {\bibinfo {author} {\bibfnamefont {I.}~\bibnamefont
  {{Wilson-Rae}}}, \bibinfo {author} {\bibfnamefont {N.}~\bibnamefont
  {Nooshi}}, \bibinfo {author} {\bibfnamefont {J.}~\bibnamefont {Dobrindt}},
  \bibinfo {author} {\bibfnamefont {T.~J.}\ \bibnamefont {Kippenberg}},\ and\
  \bibinfo {author} {\bibfnamefont {W.}~\bibnamefont {Zwerger}},\ }\href
  {https://doi.org/10.1088/1367-2630/10/9/095007} {\bibfield  {journal}
  {\bibinfo  {journal} {New J. Phys.}\ }\textbf {\bibinfo {volume} {10}},\
  \bibinfo {pages} {095007} (\bibinfo {year} {2008})}\BibitemShut {NoStop}%
\bibitem [{\citenamefont {Reitz}\ \emph {et~al.}(2020)\citenamefont {Reitz},
  \citenamefont {Sommer}, \citenamefont {Gurlek}, \citenamefont {Sandoghdar},
  \citenamefont {{Martin-Cano}},\ and\ \citenamefont {Genes}}]{Reitz2020}%
  \BibitemOpen
  \bibfield  {author} {\bibinfo {author} {\bibfnamefont {M.}~\bibnamefont
  {Reitz}}, \bibinfo {author} {\bibfnamefont {C.}~\bibnamefont {Sommer}},
  \bibinfo {author} {\bibfnamefont {B.}~\bibnamefont {Gurlek}}, \bibinfo
  {author} {\bibfnamefont {V.}~\bibnamefont {Sandoghdar}}, \bibinfo {author}
  {\bibfnamefont {D.}~\bibnamefont {{Martin-Cano}}},\ and\ \bibinfo {author}
  {\bibfnamefont {C.}~\bibnamefont {Genes}},\ }\href
  {https://doi.org/10.1103/PhysRevResearch.2.033270} {\bibfield  {journal}
  {\bibinfo  {journal} {Phys. Rev. Research}\ }\textbf {\bibinfo {volume}
  {2}},\ \bibinfo {pages} {033270} (\bibinfo {year} {2020})}\BibitemShut
  {NoStop}%
\bibitem [{\citenamefont {Negri}\ \emph {et~al.}(2002)\citenamefont {Negri},
  \citenamefont {Castiglioni}, \citenamefont {Tommasini},\ and\ \citenamefont
  {Zerbi}}]{Negri2002}%
  \BibitemOpen
  \bibfield  {author} {\bibinfo {author} {\bibfnamefont {F.}~\bibnamefont
  {Negri}}, \bibinfo {author} {\bibfnamefont {C.}~\bibnamefont {Castiglioni}},
  \bibinfo {author} {\bibfnamefont {M.}~\bibnamefont {Tommasini}},\ and\
  \bibinfo {author} {\bibfnamefont {G.}~\bibnamefont {Zerbi}},\ }\href@noop {}
  {\bibfield  {journal} {\bibinfo  {journal} {The Journal of Physical Chemistry
  A}\ }\textbf {\bibinfo {volume} {106}},\ \bibinfo {pages} {3306} (\bibinfo
  {year} {2002})}\BibitemShut {NoStop}%
\bibitem [{\citenamefont {Zirkelbach}\ \emph {et~al.}(2021)\citenamefont
  {Zirkelbach}, \citenamefont {Mirzaei}, \citenamefont {Deperasinska},
  \citenamefont {Kozankiewicz}, \citenamefont {Gurlek}, \citenamefont
  {Shkarin}, \citenamefont {Utikal}, \citenamefont {G{\"o}tzinger},\ and\
  \citenamefont {Sandoghdar}}]{Zirkelbach2021}%
  \BibitemOpen
  \bibfield  {author} {\bibinfo {author} {\bibfnamefont {J.}~\bibnamefont
  {Zirkelbach}}, \bibinfo {author} {\bibfnamefont {M.}~\bibnamefont {Mirzaei}},
  \bibinfo {author} {\bibfnamefont {I.}~\bibnamefont {Deperasinska}}, \bibinfo
  {author} {\bibfnamefont {B.}~\bibnamefont {Kozankiewicz}}, \bibinfo {author}
  {\bibfnamefont {B.}~\bibnamefont {Gurlek}}, \bibinfo {author} {\bibfnamefont
  {A.}~\bibnamefont {Shkarin}}, \bibinfo {author} {\bibfnamefont
  {T.}~\bibnamefont {Utikal}}, \bibinfo {author} {\bibfnamefont
  {S.}~\bibnamefont {G{\"o}tzinger}},\ and\ \bibinfo {author} {\bibfnamefont
  {V.}~\bibnamefont {Sandoghdar}},\ }\href@noop {} {\bibfield  {journal}
  {\bibinfo  {journal} {arXiv:2112.04806 [quant-ph]}\ } (\bibinfo {year}
  {2021})},\ \Eprint {https://arxiv.org/abs/2112.04806} {arxiv:2112.04806
  [quant-ph]} \BibitemShut {NoStop}%
\bibitem [{\citenamefont {Tommasini}\ \emph {et~al.}(2009)\citenamefont
  {Tommasini}, \citenamefont {Castiglioni},\ and\ \citenamefont
  {Zerbi}}]{Tommasini2009}%
  \BibitemOpen
  \bibfield  {author} {\bibinfo {author} {\bibfnamefont {M.}~\bibnamefont
  {Tommasini}}, \bibinfo {author} {\bibfnamefont {C.}~\bibnamefont
  {Castiglioni}},\ and\ \bibinfo {author} {\bibfnamefont {G.}~\bibnamefont
  {Zerbi}},\ }\href {https://doi.org/10.1039/b913660f} {\bibfield  {journal}
  {\bibinfo  {journal} {Phys. Chem. Chem. Phys.}\ }\textbf {\bibinfo {volume}
  {11}},\ \bibinfo {pages} {10185} (\bibinfo {year} {2009})}\BibitemShut
  {NoStop}%
\bibitem [{\citenamefont {May}\ and\ \citenamefont {K{\"u}hn}(2004)}]{May2004}%
  \BibitemOpen
  \bibfield  {author} {\bibinfo {author} {\bibfnamefont {V.}~\bibnamefont
  {May}}\ and\ \bibinfo {author} {\bibfnamefont {O.}~\bibnamefont {K{\"u}hn}},\
  }\href@noop {} {\emph {\bibinfo {title} {Charge and Energy Transfer Dynamics
  in Molecular Systems, 2nd}}}\ (\bibinfo {year} {2004})\BibitemShut {NoStop}%
\bibitem [{\citenamefont {Guthmuller}(2016)}]{Guthmuller2016}%
  \BibitemOpen
  \bibfield  {author} {\bibinfo {author} {\bibfnamefont {J.}~\bibnamefont
  {Guthmuller}},\ }\href {https://doi.org/10.1063/1.4941449} {\bibfield
  {journal} {\bibinfo  {journal} {J. Chem. Phys.}\ }\textbf {\bibinfo {volume}
  {144}},\ \bibinfo {pages} {064106} (\bibinfo {year} {2016})}\BibitemShut
  {NoStop}%
\bibitem [{sup()}]{supp3}%
  \BibitemOpen
  \href@noop {} {}\bibinfo {note} {See Supplemental Material below for the
  derivations of the minimal, the adiabatic eliminated, the collective
  oscillator and the analytical models, respectively, as well as the results of
  the Stokes line suppresions, spectra at nonzero temperatures and photon
  correlations, which includes Ref.~\cite{Garraway2011}.}\BibitemShut {Stop}%
\bibitem [{\citenamefont {Garraway}(2011)}]{Garraway2011}%
  \BibitemOpen
  \bibfield  {author} {\bibinfo {author} {\bibfnamefont {B.~M.}\ \bibnamefont
  {Garraway}},\ }\href {https://doi.org/10.1098/rsta.2010.0333} {\bibfield
  {journal} {\bibinfo  {journal} {Philosophical Transactions of the Royal
  Society A: Mathematical, Physical and Engineering Sciences}\ }\textbf
  {\bibinfo {volume} {369}},\ \bibinfo {pages} {1137} (\bibinfo {year}
  {2011})}\BibitemShut {NoStop}%
\bibitem [{\citenamefont {Gardiner}\ \emph {et~al.}(2004)\citenamefont
  {Gardiner}, \citenamefont {Zoller},\ and\ \citenamefont
  {Zoller}}]{Gardiner2004}%
  \BibitemOpen
  \bibfield  {author} {\bibinfo {author} {\bibfnamefont {C.}~\bibnamefont
  {Gardiner}}, \bibinfo {author} {\bibfnamefont {P.}~\bibnamefont {Zoller}},\
  and\ \bibinfo {author} {\bibfnamefont {P.}~\bibnamefont {Zoller}},\
  }\href@noop {} {\emph {\bibinfo {title} {Quantum {{Noise}}: {{A Handbook}} of
  {{Markovian}} and {{Non-Markovian Quantum Stochastic Methods}} with
  {{Applications}} to {{Quantum Optics}}}}}\ (\bibinfo  {publisher} {{Springer
  Science \& Business Media}},\ \bibinfo {year} {2004})\BibitemShut {NoStop}%
\bibitem [{\citenamefont {Schmidt}\ \emph {et~al.}(2021)\citenamefont
  {Schmidt}, \citenamefont {Esteban}, \citenamefont {Giedke}, \citenamefont
  {Aizpurua},\ and\ \citenamefont {{Gonz{\'a}lez-Tudela}}}]{Schmidt2021}%
  \BibitemOpen
  \bibfield  {author} {\bibinfo {author} {\bibfnamefont {M.~K.}\ \bibnamefont
  {Schmidt}}, \bibinfo {author} {\bibfnamefont {R.}~\bibnamefont {Esteban}},
  \bibinfo {author} {\bibfnamefont {G.}~\bibnamefont {Giedke}}, \bibinfo
  {author} {\bibfnamefont {J.}~\bibnamefont {Aizpurua}},\ and\ \bibinfo
  {author} {\bibfnamefont {A.}~\bibnamefont {{Gonz{\'a}lez-Tudela}}},\ }\href
  {https://doi.org/10.1088/2058-9565/abe569} {\bibfield  {journal} {\bibinfo
  {journal} {Quantum Sci. Technol.}\ }\textbf {\bibinfo {volume} {6}},\
  \bibinfo {pages} {034005} (\bibinfo {year} {2021})}\BibitemShut {NoStop}%
\bibitem [{\citenamefont {Kumar}\ and\ \citenamefont
  {Meath}(1992)}]{Kumar1992}%
  \BibitemOpen
  \bibfield  {author} {\bibinfo {author} {\bibfnamefont {A.}~\bibnamefont
  {Kumar}}\ and\ \bibinfo {author} {\bibfnamefont {W.~J.}\ \bibnamefont
  {Meath}},\ }\href {https://doi.org/10.1080/00268979200100251} {\bibfield
  {journal} {\bibinfo  {journal} {Molecular Physics}\ }\textbf {\bibinfo
  {volume} {75}},\ \bibinfo {pages} {311} (\bibinfo {year} {1992})}\BibitemShut
  {NoStop}%
\bibitem [{\citenamefont {Clear}\ \emph {et~al.}(2020)\citenamefont {Clear},
  \citenamefont {Schofield}, \citenamefont {Major}, \citenamefont
  {{Iles-Smith}}, \citenamefont {Clark},\ and\ \citenamefont
  {McCutcheon}}]{Clear2020}%
  \BibitemOpen
  \bibfield  {author} {\bibinfo {author} {\bibfnamefont {C.}~\bibnamefont
  {Clear}}, \bibinfo {author} {\bibfnamefont {R.~C.}\ \bibnamefont
  {Schofield}}, \bibinfo {author} {\bibfnamefont {K.~D.}\ \bibnamefont
  {Major}}, \bibinfo {author} {\bibfnamefont {J.}~\bibnamefont {{Iles-Smith}}},
  \bibinfo {author} {\bibfnamefont {A.~S.}\ \bibnamefont {Clark}},\ and\
  \bibinfo {author} {\bibfnamefont {D.~P.~S.}\ \bibnamefont {McCutcheon}},\
  }\href {https://doi.org/10.1103/PhysRevLett.124.153602} {\bibfield  {journal}
  {\bibinfo  {journal} {Phys. Rev. Lett.}\ }\textbf {\bibinfo {volume} {124}},\
  \bibinfo {pages} {153602} (\bibinfo {year} {2020})}\BibitemShut {NoStop}%
\bibitem [{\citenamefont {Zirkelbach}\ \emph {et~al.}(2022)\citenamefont
  {Zirkelbach}, \citenamefont {Mirzaei}, \citenamefont {Deperasi{\'n}ska},
  \citenamefont {Kozankiewicz}, \citenamefont {Gurlek}, \citenamefont
  {Shkarin}, \citenamefont {Utikal}, \citenamefont {G{\"o}tzinger},\ and\
  \citenamefont {Sandoghdar}}]{Zirkelbach2022}%
  \BibitemOpen
  \bibfield  {author} {\bibinfo {author} {\bibfnamefont {J.}~\bibnamefont
  {Zirkelbach}}, \bibinfo {author} {\bibfnamefont {M.}~\bibnamefont {Mirzaei}},
  \bibinfo {author} {\bibfnamefont {I.}~\bibnamefont {Deperasi{\'n}ska}},
  \bibinfo {author} {\bibfnamefont {B.}~\bibnamefont {Kozankiewicz}}, \bibinfo
  {author} {\bibfnamefont {B.}~\bibnamefont {Gurlek}}, \bibinfo {author}
  {\bibfnamefont {A.}~\bibnamefont {Shkarin}}, \bibinfo {author} {\bibfnamefont
  {T.}~\bibnamefont {Utikal}}, \bibinfo {author} {\bibfnamefont
  {S.}~\bibnamefont {G{\"o}tzinger}},\ and\ \bibinfo {author} {\bibfnamefont
  {V.}~\bibnamefont {Sandoghdar}},\ }\href {https://doi.org/10.1063/5.0081297}
  {\bibfield  {journal} {\bibinfo  {journal} {J. Chem. Phys.}\ }\textbf
  {\bibinfo {volume} {156}},\ \bibinfo {pages} {104301} (\bibinfo {year}
  {2022})}\BibitemShut {NoStop}%
\bibitem [{\citenamefont {Zirkelbach}\ \emph {et~al.}(2023)\citenamefont
  {Zirkelbach}, \citenamefont {Gurlek}, \citenamefont {Mirzaei}, \citenamefont
  {Shkarin}, \citenamefont {Utikal}, \citenamefont {G{\"o}tzinger},\ and\
  \citenamefont {Sandoghdar}}]{Zirkelbach2023}%
  \BibitemOpen
  \bibfield  {author} {\bibinfo {author} {\bibfnamefont {J.}~\bibnamefont
  {Zirkelbach}}, \bibinfo {author} {\bibfnamefont {B.}~\bibnamefont {Gurlek}},
  \bibinfo {author} {\bibfnamefont {M.}~\bibnamefont {Mirzaei}}, \bibinfo
  {author} {\bibfnamefont {A.}~\bibnamefont {Shkarin}}, \bibinfo {author}
  {\bibfnamefont {T.}~\bibnamefont {Utikal}}, \bibinfo {author} {\bibfnamefont
  {S.}~\bibnamefont {G{\"o}tzinger}},\ and\ \bibinfo {author} {\bibfnamefont
  {V.}~\bibnamefont {Sandoghdar}},\ }\href
  {https://doi.org/10.48550/arXiv.2302.14733} {\bibinfo {title} {Stimulated
  {{Raman}} transition in a single molecule}} (\bibinfo {year} {2023}),\
  \Eprint {https://arxiv.org/abs/2302.14733} {arxiv:2302.14733 [quant-ph]}
  \BibitemShut {NoStop}%
\bibitem [{\citenamefont {Schmidt}\ \emph {et~al.}(2017)\citenamefont
  {Schmidt}, \citenamefont {Esteban}, \citenamefont {Benz}, \citenamefont
  {Baumberg},\ and\ \citenamefont {Aizpurua}}]{Schmidt2017}%
  \BibitemOpen
  \bibfield  {author} {\bibinfo {author} {\bibfnamefont {M.~K.}\ \bibnamefont
  {Schmidt}}, \bibinfo {author} {\bibfnamefont {R.}~\bibnamefont {Esteban}},
  \bibinfo {author} {\bibfnamefont {F.}~\bibnamefont {Benz}}, \bibinfo {author}
  {\bibfnamefont {J.~J.}\ \bibnamefont {Baumberg}},\ and\ \bibinfo {author}
  {\bibfnamefont {J.}~\bibnamefont {Aizpurua}},\ }\href
  {https://doi.org/10.1039/C7FD00145B} {\bibfield  {journal} {\bibinfo
  {journal} {Faraday Discuss.}\ }\textbf {\bibinfo {volume} {205}},\ \bibinfo
  {pages} {31} (\bibinfo {year} {2017})}\BibitemShut {NoStop}%
\bibitem [{\citenamefont {{del Pino}}\ \emph {et~al.}(2018)\citenamefont {{del
  Pino}}, \citenamefont {Schr{\"o}der}, \citenamefont {Chin}, \citenamefont
  {Feist},\ and\ \citenamefont {{Garcia-Vidal}}}]{delPino2018}%
  \BibitemOpen
  \bibfield  {author} {\bibinfo {author} {\bibfnamefont {J.}~\bibnamefont {{del
  Pino}}}, \bibinfo {author} {\bibfnamefont {F.~A. Y.~N.}\ \bibnamefont
  {Schr{\"o}der}}, \bibinfo {author} {\bibfnamefont {A.~W.}\ \bibnamefont
  {Chin}}, \bibinfo {author} {\bibfnamefont {J.}~\bibnamefont {Feist}},\ and\
  \bibinfo {author} {\bibfnamefont {F.~J.}\ \bibnamefont {{Garcia-Vidal}}},\
  }\href {https://doi.org/10.1103/PhysRevLett.121.227401} {\bibfield  {journal}
  {\bibinfo  {journal} {Phys. Rev. Lett.}\ }\textbf {\bibinfo {volume} {121}},\
  \bibinfo {pages} {227401} (\bibinfo {year} {2018})}\BibitemShut {NoStop}%
\bibitem [{\citenamefont {{Fowler-Wright}}\ \emph {et~al.}(2022)\citenamefont
  {{Fowler-Wright}}, \citenamefont {Lovett},\ and\ \citenamefont
  {Keeling}}]{Fowler-Wright2022}%
  \BibitemOpen
  \bibfield  {author} {\bibinfo {author} {\bibfnamefont {P.}~\bibnamefont
  {{Fowler-Wright}}}, \bibinfo {author} {\bibfnamefont {B.~W.}\ \bibnamefont
  {Lovett}},\ and\ \bibinfo {author} {\bibfnamefont {J.}~\bibnamefont
  {Keeling}},\ }\href {https://doi.org/10.1103/PhysRevLett.129.173001}
  {\bibfield  {journal} {\bibinfo  {journal} {Phys. Rev. Lett.}\ }\textbf
  {\bibinfo {volume} {129}},\ \bibinfo {pages} {173001} (\bibinfo {year}
  {2022})}\BibitemShut {NoStop}%
\bibitem [{\citenamefont {Cygorek}\ \emph {et~al.}(2022)\citenamefont
  {Cygorek}, \citenamefont {Cosacchi}, \citenamefont {Vagov}, \citenamefont
  {Axt}, \citenamefont {Lovett}, \citenamefont {Keeling},\ and\ \citenamefont
  {Gauger}}]{Cygorek2022}%
  \BibitemOpen
  \bibfield  {author} {\bibinfo {author} {\bibfnamefont {M.}~\bibnamefont
  {Cygorek}}, \bibinfo {author} {\bibfnamefont {M.}~\bibnamefont {Cosacchi}},
  \bibinfo {author} {\bibfnamefont {A.}~\bibnamefont {Vagov}}, \bibinfo
  {author} {\bibfnamefont {V.~M.}\ \bibnamefont {Axt}}, \bibinfo {author}
  {\bibfnamefont {B.~W.}\ \bibnamefont {Lovett}}, \bibinfo {author}
  {\bibfnamefont {J.}~\bibnamefont {Keeling}},\ and\ \bibinfo {author}
  {\bibfnamefont {E.~M.}\ \bibnamefont {Gauger}},\ }\href
  {https://doi.org/10.1038/s41567-022-01544-9} {\bibfield  {journal} {\bibinfo
  {journal} {Nat. Phys.}\ }\textbf {\bibinfo {volume} {18}},\ \bibinfo {pages}
  {662} (\bibinfo {year} {2022})}\BibitemShut {NoStop}%
\bibitem [{\citenamefont {Kurucz}\ and\ \citenamefont
  {M{\o}lmer}(2010)}]{Kurucz2010}%
  \BibitemOpen
  \bibfield  {author} {\bibinfo {author} {\bibfnamefont {Z.}~\bibnamefont
  {Kurucz}}\ and\ \bibinfo {author} {\bibfnamefont {K.}~\bibnamefont
  {M{\o}lmer}},\ }\href {https://doi.org/10.1103/PhysRevA.81.032314} {\bibfield
   {journal} {\bibinfo  {journal} {Phys. Rev. A}\ }\textbf {\bibinfo {volume}
  {81}},\ \bibinfo {pages} {032314} (\bibinfo {year} {2010})}\BibitemShut
  {NoStop}%
\bibitem [{\citenamefont {Johansson}\ \emph {et~al.}(2012)\citenamefont
  {Johansson}, \citenamefont {Nation},\ and\ \citenamefont
  {Nori}}]{Johansson2012}%
  \BibitemOpen
  \bibfield  {author} {\bibinfo {author} {\bibfnamefont {{\relax
  JR}.}~\bibnamefont {Johansson}}, \bibinfo {author} {\bibfnamefont {{\relax
  PD}.}~\bibnamefont {Nation}},\ and\ \bibinfo {author} {\bibfnamefont
  {F.}~\bibnamefont {Nori}},\ }\href@noop {} {\bibfield  {journal} {\bibinfo
  {journal} {Computer Physics Communications}\ }\textbf {\bibinfo {volume}
  {183}},\ \bibinfo {pages} {1760} (\bibinfo {year} {2012})}\BibitemShut
  {NoStop}%
\bibitem [{\citenamefont {S{\'a}nchez~Mu{\~n}oz}\ \emph
  {et~al.}(2014)\citenamefont {S{\'a}nchez~Mu{\~n}oz}, \citenamefont {{del
  Valle}}, \citenamefont {Tejedor},\ and\ \citenamefont
  {Laussy}}]{SanchezMunoz2014}%
  \BibitemOpen
  \bibfield  {author} {\bibinfo {author} {\bibfnamefont {C.}~\bibnamefont
  {S{\'a}nchez~Mu{\~n}oz}}, \bibinfo {author} {\bibfnamefont {E.}~\bibnamefont
  {{del Valle}}}, \bibinfo {author} {\bibfnamefont {C.}~\bibnamefont
  {Tejedor}},\ and\ \bibinfo {author} {\bibfnamefont {F.~P.}\ \bibnamefont
  {Laussy}},\ }\href {https://doi.org/10.1103/PhysRevA.90.052111} {\bibfield
  {journal} {\bibinfo  {journal} {Phys. Rev. A}\ }\textbf {\bibinfo {volume}
  {90}},\ \bibinfo {pages} {052111} (\bibinfo {year} {2014})}\BibitemShut
  {NoStop}%
\bibitem [{\citenamefont {Klyshko}(1977)}]{Klyshko1977}%
  \BibitemOpen
  \bibfield  {author} {\bibinfo {author} {\bibfnamefont {D.~N.}\ \bibnamefont
  {Klyshko}},\ }\href {https://doi.org/10.1070/QE1977v007n06ABEH012890}
  {\bibfield  {journal} {\bibinfo  {journal} {Sov. J. Quantum Electron.}\
  }\textbf {\bibinfo {volume} {7}},\ \bibinfo {pages} {755} (\bibinfo {year}
  {1977})}\BibitemShut {NoStop}%
\bibitem [{\citenamefont {Kasperczyk}\ \emph {et~al.}(2016)\citenamefont
  {Kasperczyk}, \citenamefont {{de Aguiar J{\'u}nior}}, \citenamefont {Rabelo},
  \citenamefont {Saraiva}, \citenamefont {Santos}, \citenamefont {Novotny},\
  and\ \citenamefont {Jorio}}]{Kasperczyk2016}%
  \BibitemOpen
  \bibfield  {author} {\bibinfo {author} {\bibfnamefont {M.}~\bibnamefont
  {Kasperczyk}}, \bibinfo {author} {\bibfnamefont {F.~S.}\ \bibnamefont {{de
  Aguiar J{\'u}nior}}}, \bibinfo {author} {\bibfnamefont {C.}~\bibnamefont
  {Rabelo}}, \bibinfo {author} {\bibfnamefont {A.}~\bibnamefont {Saraiva}},
  \bibinfo {author} {\bibfnamefont {M.~F.}\ \bibnamefont {Santos}}, \bibinfo
  {author} {\bibfnamefont {L.}~\bibnamefont {Novotny}},\ and\ \bibinfo {author}
  {\bibfnamefont {A.}~\bibnamefont {Jorio}},\ }\href
  {https://doi.org/10.1103/PhysRevLett.117.243603} {\bibfield  {journal}
  {\bibinfo  {journal} {Phys. Rev. Lett.}\ }\textbf {\bibinfo {volume} {117}},\
  \bibinfo {pages} {243603} (\bibinfo {year} {2016})}\BibitemShut {NoStop}%
\bibitem [{\citenamefont {Anderson}\ \emph {et~al.}(2018)\citenamefont
  {Anderson}, \citenamefont {Tarrago~Velez}, \citenamefont {Seibold},
  \citenamefont {Flayac}, \citenamefont {Savona}, \citenamefont {Sangouard},\
  and\ \citenamefont {Galland}}]{Anderson2018}%
  \BibitemOpen
  \bibfield  {author} {\bibinfo {author} {\bibfnamefont {M.~D.}\ \bibnamefont
  {Anderson}}, \bibinfo {author} {\bibfnamefont {S.}~\bibnamefont
  {Tarrago~Velez}}, \bibinfo {author} {\bibfnamefont {K.}~\bibnamefont
  {Seibold}}, \bibinfo {author} {\bibfnamefont {H.}~\bibnamefont {Flayac}},
  \bibinfo {author} {\bibfnamefont {V.}~\bibnamefont {Savona}}, \bibinfo
  {author} {\bibfnamefont {N.}~\bibnamefont {Sangouard}},\ and\ \bibinfo
  {author} {\bibfnamefont {C.}~\bibnamefont {Galland}},\ }\href
  {https://doi.org/10.1103/PhysRevLett.120.233601} {\bibfield  {journal}
  {\bibinfo  {journal} {Phys. Rev. Lett.}\ }\textbf {\bibinfo {volume} {120}},\
  \bibinfo {pages} {233601} (\bibinfo {year} {2018})}\BibitemShut {NoStop}%
\bibitem [{\citenamefont {Toninelli}\ \emph {et~al.}(2021)\citenamefont
  {Toninelli}, \citenamefont {Gerhardt}, \citenamefont {Clark}, \citenamefont
  {{Reserbat-Plantey}}, \citenamefont {G{\"o}tzinger}, \citenamefont
  {Ristanovi{\'c}}, \citenamefont {Colautti}, \citenamefont {Lombardi},
  \citenamefont {Major}, \citenamefont {Deperasi{\'n}ska}, \citenamefont
  {Pernice}, \citenamefont {Koppens}, \citenamefont {Kozankiewicz},
  \citenamefont {Gourdon}, \citenamefont {Sandoghdar},\ and\ \citenamefont
  {Orrit}}]{Toninelli2021}%
  \BibitemOpen
  \bibfield  {author} {\bibinfo {author} {\bibfnamefont {C.}~\bibnamefont
  {Toninelli}}, \bibinfo {author} {\bibfnamefont {I.}~\bibnamefont {Gerhardt}},
  \bibinfo {author} {\bibfnamefont {A.~S.}\ \bibnamefont {Clark}}, \bibinfo
  {author} {\bibfnamefont {A.}~\bibnamefont {{Reserbat-Plantey}}}, \bibinfo
  {author} {\bibfnamefont {S.}~\bibnamefont {G{\"o}tzinger}}, \bibinfo {author}
  {\bibfnamefont {Z.}~\bibnamefont {Ristanovi{\'c}}}, \bibinfo {author}
  {\bibfnamefont {M.}~\bibnamefont {Colautti}}, \bibinfo {author}
  {\bibfnamefont {P.}~\bibnamefont {Lombardi}}, \bibinfo {author}
  {\bibfnamefont {K.~D.}\ \bibnamefont {Major}}, \bibinfo {author}
  {\bibfnamefont {I.}~\bibnamefont {Deperasi{\'n}ska}}, \bibinfo {author}
  {\bibfnamefont {W.~H.}\ \bibnamefont {Pernice}}, \bibinfo {author}
  {\bibfnamefont {F.~H.~L.}\ \bibnamefont {Koppens}}, \bibinfo {author}
  {\bibfnamefont {B.}~\bibnamefont {Kozankiewicz}}, \bibinfo {author}
  {\bibfnamefont {A.}~\bibnamefont {Gourdon}}, \bibinfo {author} {\bibfnamefont
  {V.}~\bibnamefont {Sandoghdar}},\ and\ \bibinfo {author} {\bibfnamefont
  {M.}~\bibnamefont {Orrit}},\ }\href
  {https://doi.org/10.1038/s41563-021-00987-4} {\bibfield  {journal} {\bibinfo
  {journal} {Nat. Mater.}\ ,\ \bibinfo {pages} {1}} (\bibinfo {year}
  {2021})}\BibitemShut {NoStop}%
\bibitem [{\citenamefont {Caldarola}\ \emph {et~al.}(2015)\citenamefont
  {Caldarola}, \citenamefont {Albella}, \citenamefont {Cortes}, \citenamefont
  {Rahmani}, \citenamefont {Roschuk}, \citenamefont {Grinblat}, \citenamefont
  {Oulton}, \citenamefont {Bragas},\ and\ \citenamefont
  {Maier}}]{caldarola2015}%
  \BibitemOpen
  \bibfield  {author} {\bibinfo {author} {\bibfnamefont {M.}~\bibnamefont
  {Caldarola}}, \bibinfo {author} {\bibfnamefont {P.}~\bibnamefont {Albella}},
  \bibinfo {author} {\bibfnamefont {E.}~\bibnamefont {Cortes}}, \bibinfo
  {author} {\bibfnamefont {M.}~\bibnamefont {Rahmani}}, \bibinfo {author}
  {\bibfnamefont {T.}~\bibnamefont {Roschuk}}, \bibinfo {author} {\bibfnamefont
  {G.}~\bibnamefont {Grinblat}}, \bibinfo {author} {\bibfnamefont {R.~F.}\
  \bibnamefont {Oulton}}, \bibinfo {author} {\bibfnamefont {A.~V.}\
  \bibnamefont {Bragas}},\ and\ \bibinfo {author} {\bibfnamefont {S.~A.}\
  \bibnamefont {Maier}},\ }\href {https://doi.org/10.1038/ncomms8915}
  {\bibfield  {journal} {\bibinfo  {journal} {Nature Communications}\ }\textbf
  {\bibinfo {volume} {6}},\ \bibinfo {pages} {7915} (\bibinfo {year}
  {2015})}\BibitemShut {NoStop}%
\bibitem [{\citenamefont {Albella}\ \emph {et~al.}(2013)\citenamefont
  {Albella}, \citenamefont {Poyli}, \citenamefont {Schmidt}, \citenamefont
  {Maier}, \citenamefont {Moreno}, \citenamefont {Saenz},\ and\ \citenamefont
  {Aizpurua}}]{albella2013}%
  \BibitemOpen
  \bibfield  {author} {\bibinfo {author} {\bibfnamefont {P.}~\bibnamefont
  {Albella}}, \bibinfo {author} {\bibfnamefont {M.~A.}\ \bibnamefont {Poyli}},
  \bibinfo {author} {\bibfnamefont {M.~K.}\ \bibnamefont {Schmidt}}, \bibinfo
  {author} {\bibfnamefont {S.~A.}\ \bibnamefont {Maier}}, \bibinfo {author}
  {\bibfnamefont {F.}~\bibnamefont {Moreno}}, \bibinfo {author} {\bibfnamefont
  {J.~J.}\ \bibnamefont {Saenz}},\ and\ \bibinfo {author} {\bibfnamefont
  {J.}~\bibnamefont {Aizpurua}},\ }\href {https://doi.org/10.1021/jp4027018}
  {\bibfield  {journal} {\bibinfo  {journal} {The Journal of Physical Chemistry
  C}\ }\textbf {\bibinfo {volume} {117}},\ \bibinfo {pages} {13573} (\bibinfo
  {year} {2013})}\BibitemShut {NoStop}%
\bibitem [{\citenamefont {Kuznetsov}\ \emph {et~al.}(2016)\citenamefont
  {Kuznetsov}, \citenamefont {Miroshnichenko}, \citenamefont {Brongersma},
  \citenamefont {Kivshar},\ and\ \citenamefont {Luk\'yanchuk}}]{kuznetsov2016}%
  \BibitemOpen
  \bibfield  {author} {\bibinfo {author} {\bibfnamefont {A.~I.}\ \bibnamefont
  {Kuznetsov}}, \bibinfo {author} {\bibfnamefont {A.~E.}\ \bibnamefont
  {Miroshnichenko}}, \bibinfo {author} {\bibfnamefont {M.~L.}\ \bibnamefont
  {Brongersma}}, \bibinfo {author} {\bibfnamefont {Y.~S.}\ \bibnamefont
  {Kivshar}},\ and\ \bibinfo {author} {\bibfnamefont {B.}~\bibnamefont
  {Luk\'yanchuk}},\ }\href {https://doi.org/10.1126/science.aag2472} {\bibfield
   {journal} {\bibinfo  {journal} {Science}\ }\textbf {\bibinfo {volume}
  {354}},\ \bibinfo {pages} {aag2472} (\bibinfo {year} {2016})}\BibitemShut
  {NoStop}%
\end{thebibliography}%


\begin{thebibliography}{13}%
\makeatletter
\providecommand \@ifxundefined [1]{%
 \@ifx{#1\undefined}
}%
\providecommand \@ifnum [1]{%
 \ifnum #1\expandafter \@firstoftwo
 \else \expandafter \@secondoftwo
 \fi
}%
\providecommand \@ifx [1]{%
 \ifx #1\expandafter \@firstoftwo
 \else \expandafter \@secondoftwo
 \fi
}%
\providecommand \natexlab [1]{#1}%
\providecommand \enquote  [1]{``#1''}%
\providecommand \bibnamefont  [1]{#1}%
\providecommand \bibfnamefont [1]{#1}%
\providecommand \citenamefont [1]{#1}%
\providecommand \href@noop [0]{\@secondoftwo}%
\providecommand \href [0]{\begingroup \@sanitize@url \@href}%
\providecommand \@href[1]{\@@startlink{#1}\@@href}%
\providecommand \@@href[1]{\endgroup#1\@@endlink}%
\providecommand \@sanitize@url [0]{\catcode `\\12\catcode `\$12\catcode
  `\&12\catcode `\#12\catcode `\^12\catcode `\_12\catcode `\%12\relax}%
\providecommand \@@startlink[1]{}%
\providecommand \@@endlink[0]{}%
\providecommand \url  [0]{\begingroup\@sanitize@url \@url }%
\providecommand \@url [1]{\endgroup\@href {#1}{\urlprefix }}%
\providecommand \urlprefix  [0]{URL }%
\providecommand \Eprint [0]{\href }%
\providecommand \doibase [0]{https://doi.org/}%
\providecommand \selectlanguage [0]{\@gobble}%
\providecommand \bibinfo  [0]{\@secondoftwo}%
\providecommand \bibfield  [0]{\@secondoftwo}%
\providecommand \translation [1]{[#1]}%
\providecommand \BibitemOpen [0]{}%
\providecommand \bibitemStop [0]{}%
\providecommand \bibitemNoStop [0]{.\EOS\space}%
\providecommand \EOS [0]{\spacefactor3000\relax}%
\providecommand \BibitemShut  [1]{\csname bibitem#1\endcsname}%
\let\auto@bib@innerbib\@empty
\bibitem [{\citenamefont {May}\ and\ \citenamefont {K{\"u}hn}(2004)}]{May2004}%
  \BibitemOpen
  \bibfield  {author} {\bibinfo {author} {\bibfnamefont {V.}~\bibnamefont
  {May}}\ and\ \bibinfo {author} {\bibfnamefont {O.}~\bibnamefont {K{\"u}hn}},\
  }\href@noop {} {\emph {\bibinfo {title} {Charge and Energy Transfer Dynamics
  in Molecular Systems, 2nd}}}\ (\bibinfo {year} {2004})\BibitemShut {NoStop}%
\bibitem [{\citenamefont {Albrecht}\ and\ \citenamefont
  {Creighton}(1977)}]{Albrecht1977}%
  \BibitemOpen
  \bibfield  {author} {\bibinfo {author} {\bibfnamefont {M.~G.}\ \bibnamefont
  {Albrecht}}\ and\ \bibinfo {author} {\bibfnamefont {J.~A.}\ \bibnamefont
  {Creighton}},\ }\href@noop {} {\bibfield  {journal} {\bibinfo  {journal} {J.
  Am. Chem. Soc.}\ }\textbf {\bibinfo {volume} {99}},\ \bibinfo {pages} {5215}
  (\bibinfo {year} {1977})}\BibitemShut {NoStop}%
\bibitem [{\citenamefont {Guthmuller}(2016)}]{Guthmuller2016}%
  \BibitemOpen
  \bibfield  {author} {\bibinfo {author} {\bibfnamefont {J.}~\bibnamefont
  {Guthmuller}},\ }\href {https://doi.org/10.1063/1.4941449} {\bibfield
  {journal} {\bibinfo  {journal} {J. Chem. Phys.}\ }\textbf {\bibinfo {volume}
  {144}},\ \bibinfo {pages} {064106} (\bibinfo {year} {2016})}\BibitemShut
  {NoStop}%
\bibitem [{\citenamefont {Reitz}\ \emph {et~al.}(2020)\citenamefont {Reitz},
  \citenamefont {Sommer}, \citenamefont {Gurlek}, \citenamefont {Sandoghdar},
  \citenamefont {{Martin-Cano}},\ and\ \citenamefont {Genes}}]{Reitz2020}%
  \BibitemOpen
  \bibfield  {author} {\bibinfo {author} {\bibfnamefont {M.}~\bibnamefont
  {Reitz}}, \bibinfo {author} {\bibfnamefont {C.}~\bibnamefont {Sommer}},
  \bibinfo {author} {\bibfnamefont {B.}~\bibnamefont {Gurlek}}, \bibinfo
  {author} {\bibfnamefont {V.}~\bibnamefont {Sandoghdar}}, \bibinfo {author}
  {\bibfnamefont {D.}~\bibnamefont {{Martin-Cano}}},\ and\ \bibinfo {author}
  {\bibfnamefont {C.}~\bibnamefont {Genes}},\ }\href
  {https://doi.org/10.1103/PhysRevResearch.2.033270} {\bibfield  {journal}
  {\bibinfo  {journal} {Phys. Rev. Research}\ }\textbf {\bibinfo {volume}
  {2}},\ \bibinfo {pages} {033270} (\bibinfo {year} {2020})}\BibitemShut
  {NoStop}%
\bibitem [{\citenamefont {{del Pino}}\ \emph {et~al.}(2018)\citenamefont {{del
  Pino}}, \citenamefont {Schr{\"o}der}, \citenamefont {Chin}, \citenamefont
  {Feist},\ and\ \citenamefont {{Garcia-Vidal}}}]{delPino2018}%
  \BibitemOpen
  \bibfield  {author} {\bibinfo {author} {\bibfnamefont {J.}~\bibnamefont {{del
  Pino}}}, \bibinfo {author} {\bibfnamefont {F.~A. Y.~N.}\ \bibnamefont
  {Schr{\"o}der}}, \bibinfo {author} {\bibfnamefont {A.~W.}\ \bibnamefont
  {Chin}}, \bibinfo {author} {\bibfnamefont {J.}~\bibnamefont {Feist}},\ and\
  \bibinfo {author} {\bibfnamefont {F.~J.}\ \bibnamefont {{Garcia-Vidal}}},\
  }\href {https://doi.org/10.1103/PhysRevLett.121.227401} {\bibfield  {journal}
  {\bibinfo  {journal} {Phys. Rev. Lett.}\ }\textbf {\bibinfo {volume} {121}},\
  \bibinfo {pages} {227401} (\bibinfo {year} {2018})}\BibitemShut {NoStop}%
\bibitem [{\citenamefont {{Fowler-Wright}}\ \emph {et~al.}(2022)\citenamefont
  {{Fowler-Wright}}, \citenamefont {Lovett},\ and\ \citenamefont
  {Keeling}}]{Fowler-Wright2022}%
  \BibitemOpen
  \bibfield  {author} {\bibinfo {author} {\bibfnamefont {P.}~\bibnamefont
  {{Fowler-Wright}}}, \bibinfo {author} {\bibfnamefont {B.~W.}\ \bibnamefont
  {Lovett}},\ and\ \bibinfo {author} {\bibfnamefont {J.}~\bibnamefont
  {Keeling}},\ }\href {https://doi.org/10.1103/PhysRevLett.129.173001}
  {\bibfield  {journal} {\bibinfo  {journal} {Phys. Rev. Lett.}\ }\textbf
  {\bibinfo {volume} {129}},\ \bibinfo {pages} {173001} (\bibinfo {year}
  {2022})}\BibitemShut {NoStop}%
\bibitem [{\citenamefont {Cygorek}\ \emph {et~al.}(2022)\citenamefont
  {Cygorek}, \citenamefont {Cosacchi}, \citenamefont {Vagov}, \citenamefont
  {Axt}, \citenamefont {Lovett}, \citenamefont {Keeling},\ and\ \citenamefont
  {Gauger}}]{Cygorek2022}%
  \BibitemOpen
  \bibfield  {author} {\bibinfo {author} {\bibfnamefont {M.}~\bibnamefont
  {Cygorek}}, \bibinfo {author} {\bibfnamefont {M.}~\bibnamefont {Cosacchi}},
  \bibinfo {author} {\bibfnamefont {A.}~\bibnamefont {Vagov}}, \bibinfo
  {author} {\bibfnamefont {V.~M.}\ \bibnamefont {Axt}}, \bibinfo {author}
  {\bibfnamefont {B.~W.}\ \bibnamefont {Lovett}}, \bibinfo {author}
  {\bibfnamefont {J.}~\bibnamefont {Keeling}},\ and\ \bibinfo {author}
  {\bibfnamefont {E.~M.}\ \bibnamefont {Gauger}},\ }\href
  {https://doi.org/10.1038/s41567-022-01544-9} {\bibfield  {journal} {\bibinfo
  {journal} {Nat. Phys.}\ }\textbf {\bibinfo {volume} {18}},\ \bibinfo {pages}
  {662} (\bibinfo {year} {2022})}\BibitemShut {NoStop}%
\bibitem [{\citenamefont {Kurucz}\ and\ \citenamefont
  {M{\o}lmer}(2010)}]{Kurucz2010}%
  \BibitemOpen
  \bibfield  {author} {\bibinfo {author} {\bibfnamefont {Z.}~\bibnamefont
  {Kurucz}}\ and\ \bibinfo {author} {\bibfnamefont {K.}~\bibnamefont
  {M{\o}lmer}},\ }\href {https://doi.org/10.1103/PhysRevA.81.032314} {\bibfield
   {journal} {\bibinfo  {journal} {Phys. Rev. A}\ }\textbf {\bibinfo {volume}
  {81}},\ \bibinfo {pages} {032314} (\bibinfo {year} {2010})}\BibitemShut
  {NoStop}%
\bibitem [{\citenamefont {Garraway}(2011)}]{Garraway2011}%
  \BibitemOpen
  \bibfield  {author} {\bibinfo {author} {\bibfnamefont {B.~M.}\ \bibnamefont
  {Garraway}},\ }\href {https://doi.org/10.1098/rsta.2010.0333} {\bibfield
  {journal} {\bibinfo  {journal} {Philosophical Transactions of the Royal
  Society A: Mathematical, Physical and Engineering Sciences}\ }\textbf
  {\bibinfo {volume} {369}},\ \bibinfo {pages} {1137} (\bibinfo {year}
  {2011})}\BibitemShut {NoStop}%
\bibitem [{\citenamefont {Neuman}\ \emph {et~al.}(2019)\citenamefont {Neuman},
  \citenamefont {Esteban}, \citenamefont {Giedke}, \citenamefont {Schmidt},\
  and\ \citenamefont {Aizpurua}}]{Neuman2019}%
  \BibitemOpen
  \bibfield  {author} {\bibinfo {author} {\bibfnamefont {T.}~\bibnamefont
  {Neuman}}, \bibinfo {author} {\bibfnamefont {R.}~\bibnamefont {Esteban}},
  \bibinfo {author} {\bibfnamefont {G.}~\bibnamefont {Giedke}}, \bibinfo
  {author} {\bibfnamefont {M.~K.}\ \bibnamefont {Schmidt}},\ and\ \bibinfo
  {author} {\bibfnamefont {J.}~\bibnamefont {Aizpurua}},\ }\href
  {https://doi.org/10.1103/PhysRevA.100.043422} {\bibfield  {journal} {\bibinfo
   {journal} {Phys. Rev. A}\ }\textbf {\bibinfo {volume} {100}},\ \bibinfo
  {pages} {043422} (\bibinfo {year} {2019})}\BibitemShut {NoStop}%
\bibitem [{\citenamefont {Roelli}\ \emph {et~al.}(2016)\citenamefont {Roelli},
  \citenamefont {Galland}, \citenamefont {Piro},\ and\ \citenamefont
  {Kippenberg}}]{Roelli2016}%
  \BibitemOpen
  \bibfield  {author} {\bibinfo {author} {\bibfnamefont {P.}~\bibnamefont
  {Roelli}}, \bibinfo {author} {\bibfnamefont {C.}~\bibnamefont {Galland}},
  \bibinfo {author} {\bibfnamefont {N.}~\bibnamefont {Piro}},\ and\ \bibinfo
  {author} {\bibfnamefont {T.~J.}\ \bibnamefont {Kippenberg}},\ }\href
  {https://doi.org/10.1038/nnano.2015.264} {\bibfield  {journal} {\bibinfo
  {journal} {Nature Nanotechnology}\ }\textbf {\bibinfo {volume} {11}},\
  \bibinfo {pages} {164} (\bibinfo {year} {2016})}\BibitemShut {NoStop}%
\bibitem [{\citenamefont {{Wilson-Rae}}\ \emph {et~al.}(2008)\citenamefont
  {{Wilson-Rae}}, \citenamefont {Nooshi}, \citenamefont {Dobrindt},
  \citenamefont {Kippenberg},\ and\ \citenamefont {Zwerger}}]{Wilson-Rae2008}%
  \BibitemOpen
  \bibfield  {author} {\bibinfo {author} {\bibfnamefont {I.}~\bibnamefont
  {{Wilson-Rae}}}, \bibinfo {author} {\bibfnamefont {N.}~\bibnamefont
  {Nooshi}}, \bibinfo {author} {\bibfnamefont {J.}~\bibnamefont {Dobrindt}},
  \bibinfo {author} {\bibfnamefont {T.~J.}\ \bibnamefont {Kippenberg}},\ and\
  \bibinfo {author} {\bibfnamefont {W.}~\bibnamefont {Zwerger}},\ }\href
  {https://doi.org/10.1088/1367-2630/10/9/095007} {\bibfield  {journal}
  {\bibinfo  {journal} {New J. Phys.}\ }\textbf {\bibinfo {volume} {10}},\
  \bibinfo {pages} {095007} (\bibinfo {year} {2008})}\BibitemShut {NoStop}%
\bibitem [{\citenamefont {S{\'a}nchez~Mu{\~n}oz}\ \emph
  {et~al.}(2014)\citenamefont {S{\'a}nchez~Mu{\~n}oz}, \citenamefont {{del
  Valle}}, \citenamefont {Tejedor},\ and\ \citenamefont
  {Laussy}}]{SanchezMunoz2014}%
  \BibitemOpen
  \bibfield  {author} {\bibinfo {author} {\bibfnamefont {C.}~\bibnamefont
  {S{\'a}nchez~Mu{\~n}oz}}, \bibinfo {author} {\bibfnamefont {E.}~\bibnamefont
  {{del Valle}}}, \bibinfo {author} {\bibfnamefont {C.}~\bibnamefont
  {Tejedor}},\ and\ \bibinfo {author} {\bibfnamefont {F.~P.}\ \bibnamefont
  {Laussy}},\ }\href {https://doi.org/10.1103/PhysRevA.90.052111} {\bibfield
  {journal} {\bibinfo  {journal} {Phys. Rev. A}\ }\textbf {\bibinfo {volume}
  {90}},\ \bibinfo {pages} {052111} (\bibinfo {year} {2014})}\BibitemShut
  {NoStop}%
\end{thebibliography}%

\end{document}


\renewcommand{\figurename}{Fig.}
\renewcommand{\thefigure}{S\arabic{figure}}
\title{Supplementary Material: Coherent electron-vibron interactions in Surface-Enhanced Raman Scattering (SERS)}

\author{Miguel A. Martinez-Garcia}
\author{Diego Martin-Cano}
\affiliation{Departamento de F\'{i}s\'{i}ca Te\'{o}rica de la Materia Condensada and Condensed Matter Physics
Center (IFIMAC), Universidad Aut\'{o}noma de Madrid, E28049 Madrid, Spain}

\date{\today}

\maketitle
\section{Fixed frame and rotating frame Hamiltonians for the  minimal molecular model}
The Hamiltonian for our system for a single vibrational mode is~\cite{May2004,Albrecht1977,Guthmuller2016,Reitz2020}
\begin{align}
	\label{eqn:Hamiltonian1}
	\tag{S1}
	\tilde{H}&= \sum^N_{i=1} \hbar \omega_i {\xi}_i ^\dagger{\xi}_i+ \hbar \omega_\mathrm{c}{a} ^\dagger{a}+ \hbar \omega_v \hat{v}^\dagger \hat{v}+\hbar2\Omega\mathrm{cos}\left[{\omega_Lt}\right]({a}^\dagger + {a})\\  \nonumber
	&+\sum^N_{i=1}\hbar g_{0,i} {\xi}_i^\dagger {\xi}_i (\hat{v}^\dagger +\hat{v})+\sum^N_{i=1} \hbar g_i ({\xi}_i^\dagger+{\xi}_i)( {a}^\dagger +{a}),
\end{align}
where $\xi_i$  ($\xi_i^\dagger$) is the lowering (raising) operator for the the i-th electronic transition of the molecule having a frequency $\omega_i$.  $v$ ($v^\dagger$) and $a$ ($a^\dagger$) are the annihilation (creation) operators for the vibration mode of the molecule and the cavity mode, respectively. $\hbar\Omega$ is the pumping amplitude over the plasmon mode, being $\hbar\omega_L$ the laser frequency.  $\hbar g_i$ accounts for the amplitude of the Rabi interaction between the plasmon mode and the i-th electronic transition, whereas $\hbar g_{0,i}$ is the coupling between the electron and the vibron through a displaced non-linear term. The commutation relations for the operators are $[a,a^\dagger]=[v,v^\dagger]=1$ and $[\xi_i,\xi_j^\dagger]=\ket{e_i}\bra{e_j}-\delta_{ij}\ket{g}\bra{g}$, the latter differing from a standard two-level system as a result of the multiple electronic transitions sharing a common ground state.

In order to avoid the time dependence, we rewrite the Hamiltonian in the laser rotating frame for both, cavity and electronic transitions, i.e., to apply the transformation $H=U\tilde{H}U^{-1}+i(\partial_tU)U^{-1}$, where $U=e^{i\omega_L\left(a^\dagger a+\sum^N_{i=1}{\xi}_i ^\dagger{\xi}_i\right)t}$ is an unitary operator.
Considering $\left[B,\left[B,A\right]\right]=A\rightarrow e^{i\phi B}Ae^{-i\phi B}=A\mathrm{cos}\phi+i\left[B,A\right]\mathrm{sin}\phi$, we proceed and obtain the rotating frame Hamiltonian
\begin{align}
	\tag{S2}
	\label{eqn:Hamiltonian}
	H&= \sum^N_{i=1} \hbar \Delta_i \hat{\xi}_i ^\dagger\hat{\xi}_i+ \hbar \Delta_\mathrm{c} \hat{a} ^\dagger\hat{a}+ \hbar \omega_v \hat{v}^\dagger \hat{v}+\hbar\Omega(\hat{a}^\dagger + \hat{a})\\  \nonumber
	&+\sum^N_{i=1}\hbar g_{0,i} \hat{\xi}_i^\dagger \hat{\xi}_i (\hat{v}^\dagger +\hat{v})+\sum^N_{i=1} \hbar g_i (\hat{\xi}_i^\dagger \hat{a} +  \hat{\xi}_i \hat{a}^\dagger),
\end{align}
where $\Delta_i=\omega_i-\omega_L$ and $\Delta_c=\omega_c-\omega_L$. Note that we have performed a rotating-wave approximation (RWA), assuming that both, the pumping $\Omega$ and the coupling $g$ are small compared with the cavity and the transition frequencies ($\Omega\sim g\ll \omega_c\sim\omega_r$).
\section{Adiabatic elimination for deriving the optomechanical coupling corresponding to N electronic transitions}
Following the Hamiltonian $H$ dynamics, we derive the Heisenberg equation for the i-th electronic transition
\begin{align}
	\tag{S3}
	\label{eqn:HEqAE}
	\dot{\hat{\xi_i}}&=-i\Delta_i\hat{\xi}_i-ig_i\hat{a}[1-\sum^N_{j=1}(1+\frac{g_j}{g_i})\hat{\xi_j}^\dagger\hat{\xi_i}]-ig_{0,i}\hat{\xi}_i \left(\hat{v}^\dagger+\hat{v}\right),
\end{align}
The far off-resonant electronic transitions exhibit fast dynamics and low excitations that allows for the application of an adiabatic elimination to retain only their projected effects on the slower subsystems dynamics and the leading linear term of the cavity. Therefore, we consider $\dot{\xi_i}=0$, obtaining the following approximated expression neglecting the nonlinear terms,
\begin{align}
	\tag{S4}
	\label{eqn:AE2}
	\hat{\xi_i}&\approx-\frac{g_i}{\Delta_i}\hat{a}.
\end{align}
Replacing in $H$, we obtain

\begin{align}
	\tag{S5}
	\label{eqn:AE3}
	H_{\mathrm{ae}}=\hbar \Delta^{\prime}_c \hat{a} ^\dagger\hat{a}+ \hbar\omega_v \hat{v}^\dagger \hat{v}+\hbar\Omega(\hat{a}^\dagger + \hat{a})+\hbar g_{\mathrm{om}} \hat{a}^\dagger \hat{a} (\hat{v}^\dagger + \hat{v})
\end{align}
where
\begin{align}
	\tag{S6}
	\label{eqn:AEp}
	\Delta^{\prime}_c&= \Delta_c-\sum^N_{i=1}\frac{g_i^2}{\Delta_i} \hspace{3cm}
	g_\mathrm{om}=\sum^N_{i=1} g_{0,i}  \frac{g_i^2}{\Delta_i^2}
\end{align}
\section{Collective boson operators}
The Hamiltonian $H$ is largely complex and requires highly demanding numerical methods~\cite{delPino2018,Fowler-Wright2022,Cygorek2022}  or further approximations to solve it. In this line, considering the similar coupling parameters for ultraviolet optical transitions, so $g_{0,i}\approx g_0$ and $\omega_{i}\approx \omega_b$ and the low excitation of highly detuned electronic states, we can rewrite the molecular Hamiltonian in terms of a bosonic operators \cite{Kurucz2010} and a collective bright  operator with annihilation operator, $\hat{b}= g \sum_i{\xi}_i/\sqrt{N}$, which leads to a superradiant coupling to the cavity mode.  Regarding the electron-vibron interactions, the predominant term can be rewritten in terms of the populations of the bright state due to the negligible contributions of the dark electronic states  for large off-resonant drivings, which leads to the following Hamiltonian
\begin{align}
	\tag{S7}
	\label{eqn:HamiltonianHolstein}
	H_{\mathrm{B}}&\approx\hbar \Delta_b \hat{b} ^\dagger\hat{b}+ \hbar \Delta_\mathrm{c} \hat{a} ^\dagger\hat{a}+ \hbar \omega_v \hat{v}^\dagger \hat{v}+\hbar\Omega(\hat{a}^\dagger + \hat{a})\\  \nonumber
	&+ \hbar g_0 \hat{b}^\dagger \hat{b} (\hat{v}^\dagger + \hat{v})+\sqrt{N} \hbar g (\hat{b}^\dagger \hat{a} +  \hat{b} \hat{a}^\dagger),
\end{align}
where $b$ ($b^\dagger$)  is the annihilation (creation) operator for the boson that contains all the electronic off-resonant contributions in this approximation. We notice that in the same level of approximations, a similar Hamiltonian can be analogously derived in terms of N two level systems within the Holstein-Primakoff approximation \cite{Garraway2011}.

\section{Analytical description of the anti-Stokes Raman line for the minimal model considering both, near and off-resonant transitions}
In what follows, we will consider the cavity displaced to the vacuum through $\mathcal{D}_{\alpha_s}=e^{{\alpha_s} a^\dagger-{\alpha_s}^* a}$, where ${\alpha_s}=-\Omega/(\Delta_c-i(k/2))$. Thus, the displaced cavity Hamiltonian is written

\begin{align}
	\tag{S8}
	\label{eqn:AdiabaticEliminationHamiltonian}
	H^{\alpha_s}_{\mathrm{om+ res}}&= \hbar\Delta^\prime_c \hat a^\dagger \hat a+\hbar\omega_v\hat v^\dagger \hat v+\hbar \Delta_r \hat{\xi}_r ^\dagger\hat{\xi}_r+ \hbar g_{0,r} \hat{\xi}_r^\dagger \hat{\xi}_r (\hat{v}^\dagger + \hat{v})\\ \nonumber &+ \hbar g_r (\hat{\xi}_r^\dagger  \hat{a}+ \hat{\xi}_r \hat{a}^\dagger )+ \hbar g_r ({\alpha_s}\hat{\xi}_r^\dagger + {\alpha_s}^*\hat{\xi}_r)+\hbar g_\mathrm{om} ({\alpha_s}^*\hat a+{\alpha_s} \hat a^\dagger+\hat a^\dagger \hat a)(\hat{v}^\dagger + \hat{v}) .
\end{align}

The Stokes  and anti-Stokes lines of the emission spectrum are associated to the fluctuating part of the plasmon dynamics, thus, we  linearize the Hamiltonian $H^{\alpha_s}_\mathrm{om+res}$ using  $\hat a=\mean{\hat a}+\delta \hat a$, and then, derive the following Heisenberg-Langevin equation,
\begin{align}
	\tag{S9}
	\label{eqn:HEq}
	\dot{\delta \hat a}&=\left(-i\Delta^\prime_c-\frac{\kappa}{2}\right)\delta \hat a-ig\hat\xi_r-ig_{\mathrm{om}}{\alpha_s} \left(\hat v^\dagger+\hat v\right)-ig_{\mathrm{om}}\delta \hat a \left(\hat v^\dagger+\hat v\right)+\mathcal{T}
\end{align}
where we have added a noise term $\mathcal{T}$ in order to preserve the commutation relations. We can identify $-ig\hat\xi_r$ and $-ig_{\mathrm{om}}{\alpha_s} \left(\hat v^\dagger+\hat v\right)$   as the terms containing, besides others, the Raman dynamics corresponding to the near and off-resonant transitions, respectively. We neglect the third term, since $\delta \hat a\ll {\alpha_s}$.
We assume that, the solution to this equation, i.e., the plasmonic fluctuating part, can be decomposed as follows,
\begin{align}
	\tag{S10}
	\label{eqn:HEq}
	\delta \hat a&\approx \delta \hat a^\mathrm{off}_\mathrm{R}+\delta \hat a^\mathrm{res} +\tilde{\mathcal{T}},
\end{align}
where $\delta \hat a^\mathrm{off}_\mathrm{R}=-ig_{\mathrm{om}}\int_{-\infty}^{t}e^{\left(-i\Delta^\prime_c-\frac{\kappa}{2}\right)\left(t-t^\prime\right)}{\alpha_s}\left(\hat v^\dagger+\hat v\right)dt^\prime$ and $\delta \hat a^\mathrm{res}=-ig\int_{-\infty}^{t}e^{\left(-i\Delta^\prime_c-\frac{\kappa}{2}\right)\left(t-t^\prime\right)}\xi dt^\prime $, where $\tilde{\mathcal{T}}$ represents the rest of noise terms. Following reference~\cite{Neuman2019}, we drop the noise
terms $\tilde{\mathcal{T}}$ in the
following derivation of the Raman contributions.

Our derivation starts with the calculation of the off-resonant portion of the Raman scattering, which is encompassed in the optomechanical interaction
\begin{align}
	\tag{S11}
	\label{eqn:RamanOff}
	\delta \hat a^{\mathrm{off}}_\mathrm{R}&=-ig_{\mathrm{om}}{\alpha_s}\int_{-\infty}^{t}e^{\left(-i\Delta^\prime_c-\frac{\kappa}{2}\right)\left(t-t^\prime\right)}\left(\hat v^\dagger+\hat v\right)dt^\prime\\ \nonumber
	&=-ig_{\mathrm{om}}{\alpha_s}\left[e^{i\omega_v t}\tilde{v}^\dagger\int_{-\infty}^{t}e^{\left[-i(\Delta^\prime_c+\omega_v)-\frac{\kappa}{2}\right]\left(t-t^\prime\right)}dt^\prime+e^{-i\omega_v t}\tilde{v}\int_{-\infty}^{t}e^{\left[-i(\Delta^\prime_c-\omega_v)-\frac{\kappa}{2}\right]\left(t-t^\prime\right)}dt^\prime\right]\\ \nonumber
	&=C^\mathrm{off}_{r_{S}}e^{i\omega_vt}\tilde{v}^\dagger+C^\mathrm{off}_{r_{aS}}e^{-i\omega_vt}\tilde{v},
\end{align}
where in the second step, we expressed the vibrons in the interaction picture to facilitate the application of the Markov approximation in order to solve the integral
\begin{align}
	\tag{S12}
	\label{eqn:Coefficientsoff}
	C^\mathrm{off}_{r_{S}}&= \frac{i g_{\mathrm{om}} {\alpha_s}}{\left[ i(\Delta^\prime_c+\omega_v)+\frac{\kappa}{2} \right]} ,\hspace{3cm}
	C^\mathrm{off}_{r_{aS}}= \frac{i g_{\mathrm{om}} {\alpha_s}}{\left[ i(\Delta^\prime_c-\omega_v)+\frac{\kappa}{2} \right]}.
\end{align}

The plasmon in the weak-coupling approximation can be understood as a reservoir damping the transition with a rate $\Gamma_\mathrm{eff}=\frac{g²\kappa}{(\kappa/2)²+(\Delta_r-\Delta^\prime_c)² }$ via the Purcell effect. After linearizing the transition creation operator  ($\hat\xi_r=\langle\hat\xi_r\rangle+\delta\hat\xi$), where $\langle\hat\xi_r\rangle\approx-ig_r{\alpha_s}_s/[i\Delta_r+\Gamma_\mathrm{eff}/2]$, and assuming the separation $\delta \hat a^\mathrm{res}=\delta \hat a_R^\mathrm{res}+...$, the Raman dynamics can be characterized, for $\delta\hat\xi_r\ll 1$, by~\cite{Neuman2019} $\dot{\delta\hat\xi}_\mathrm{r\,R}=(i\Delta_r-\Gamma_\mathrm{eff})\delta\hat\xi_\mathrm{r\,R}-ig_0\langle\hat\xi\rangle(v^\dagger+v)$.  Thus, we compute the  near-resonant contribution $\delta \hat a_R^\mathrm{res}$,
\begin{align}
	\tag{S13}
	\label{eqn:RamanNear}
	\delta \hat a^{\mathrm{res}}_\mathrm{R}&=-ig\int_{-\infty}^{t}e^{\left(-i\Delta^\prime_c-\frac{\kappa}{2}\right)\left(t-t^\prime\right)}\delta\hat\xi_\mathrm{R}dt^\prime\\ \nonumber &=-ig\int_{-\infty}^{t}e^{\left(-i\Delta^\prime_c-\frac{\kappa}{2}\right)\left(t-t^\prime\right)}\left[-ig_{0,r}\langle\hat\xi_r\rangle\int_{-\infty}^{t^\prime}e^{\left(-i\Delta_r-\frac{\Gamma_\mathrm{eff}}{2}\right)\left(t^\prime-t^{\prime\prime}\right)}\left(\hat v^\dagger+\hat v\right)dt^{\prime\prime}\right]dt^\prime
	\\ \nonumber
	&=C^\mathrm{res}_{r_{S}}e^{i\omega_vt}\tilde{v}^\dagger+C^\mathrm{res}_{r_{aS}}e^{-i\omega_vt}\tilde{v},
\end{align}
where
\begin{align}
	\tag{S14}
	\label{eqn:Coefficientsdd}
	C^\mathrm{res}_{r_{S}}&= \frac{-g g_{0,r} \mean{\hat\xi_r}}{\left [ i(\Delta_r+\omega_v)+\frac{\Gamma_\mathrm{eff}}{2} \right ]  \left [ i(\Delta_c^\prime+\omega_v)+\frac{\kappa}{2} \right ]}, \hspace{3cm}
	C^\mathrm{res}_{r_{aS}}= \frac{-g g_{0,r} \mean{\hat\xi_r}}{\left [ i(\Delta_r-\omega_v)+\frac{\Gamma_\mathrm{eff}}{2} \right ]  \left [ i(\Delta_c^\prime-\omega_v)+\frac{\kappa}{2} \right ]}.
\end{align}

Our objective is to calculate the emission spectrum, which is expressed as $S^e(\omega)=2\Re{\int_{-\infty}^{t}\mean{\hat a^\dagger(0) \hat a(\tau)}e^{i(\omega-\omega_L)\tau}d\tau}$. Once defined the total Raman contribution to the plasmonic fluctuating part $\delta \hat a_R=\delta \hat a^\mathrm{off}_R+\delta \hat a^\mathrm{res}_R$ , we compute the Raman constituent of the spectrum:
\begin{align}
	\tag{S15}
	\label{eqn:spectrum}
	S^\mathrm{theo} _\mathrm{R}(\omega)&=2\Re{\int_{-\infty}^{t}\mean{\delta \hat a_R^\dagger(0)\delta \hat a_R(\tau)}e^{i(\omega-\omega_L)\tau}d\tau} \\ \nonumber
	&=2\Re{\int_{-\infty}^{t}\mean{\left(\delta \hat a^\mathrm{off}_R+\delta \hat a^\mathrm{res}_R\right)^\dagger(0)\left(\delta \hat a^\mathrm{off}_R+\delta \hat a^\mathrm{res}_R\right)(\tau)}e^{i(\omega-\omega_L)\tau}d\tau} \\ \nonumber
	&=S^\mathrm{theo} _\mathrm{R,S}(\omega)+S^\mathrm{theo} _\mathrm{R,aS}(\omega),
\end{align}
where the Stokes and anti-Stokes lines of the spectra are given by,
\begin{align}
	\tag{S16}
	\label{SS}
	S^\mathrm{theo} _\mathrm{R,S}(\omega)&=\left|C^\mathrm{off}_{r_{S}}+C^\mathrm{res}_{r_{S}}\right|^2\left|C^v_{r_{S}}\left(\langle v^\dagger v\rangle_\mathrm{om+res},\omega\right)\right|^2, 
\end{align}
\begin{align}
	\tag{S17}
	S^\mathrm{theo} _\mathrm{R,aS}(\omega)&=\left|C^\mathrm{off}_{r_{aS}}+C^\mathrm{res}_{r_{aS}}\right|^2\left|C^v_{r_{aS}}\left(\langle v^\dagger v\rangle_\mathrm{om+res},\omega\right)\right|^2,
	\label{SaS}
\end{align}
and for the vibronic part, we have,
\begin{align}
	\tag{S18}
	\label{eqn:spectrum}
	|C^v_{r_{S}}\left(\langle v^\dagger v\rangle,\omega\right)|^2&=2\Re{\int_{-\infty}^{t}\mean{\tilde{v}^\dagger(0)\tilde{v}(\tau)}e^{-i(\omega_L-\omega_v)\tau}e^{i\omega\tau}d\tau}
	\approx\frac{\gamma_v \left(1+\mean{v^\dagger v}\right)}{(\omega_L-\omega_v-\omega)^2+(\frac{\gamma_v}{2})^2}, \\ \nonumber 
	|C^v_{r_{aS}}\left(\langle v^\dagger v\rangle,\omega\right)|^2&=2\Re{\int_{-\infty}^{t}\mean{\tilde{v}^\dagger(0)\tilde{v}(\tau)}e^{-i(\omega_L+\omega_v)\tau}e^{i\omega\tau}d\tau} 
	\approx\frac{\gamma_v \mean{v^\dagger v}}{(\omega_L+\omega_v-\omega)^2+(\frac{\gamma_v}{2})^2},
\end{align}
where we assumed that the optomechanical effective damping is negligible compared with the natural vibron decay. It is straightforward to show that the spectra anti-Stokes spectrum contribution for the near-resonant and the optomechanical models write 
$S^\mathrm{res,theo} _\mathrm{R,aS}(\omega)=\left|C^\mathrm{res}_{R_{aS}}\right|^2\left|C^v_{R_{aS}}\left(\langle v^\dagger v\rangle_\mathrm{res},\omega\right)\right|^2$ and $S^\mathrm{om,theo} _\mathrm{R,aS}(\omega)=\left|C^\mathrm{off}_{R_{aS}}\right|^2\left|C^v_{R_{aS}}\left(\langle v^\dagger v\rangle_\mathrm{om},\omega\right)\right|^2$, respectively, and similarly for the Stokes lines. Notice that the sums of off-resonant and near resonant Raman coefficients in eqs~\ref{SS}-\ref{SaS} are responsible for the cooperative enhancement and suppression between near resonant and off-resonant transitions in the Raman peaks for both Stokes and anti-Stokes processes.
\section{Stokes line suppression}
\begin{figure}[h]
	\begin{center}
		\includegraphics[width=8.6cm]{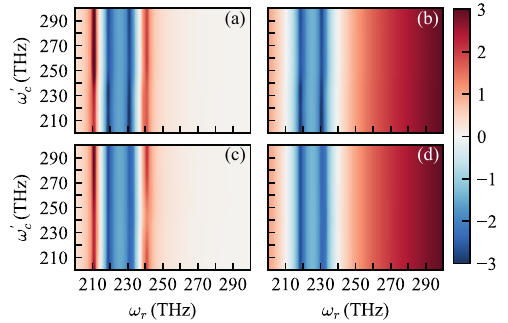}
	\end{center}
	\vspace{-0.2cm}
	\caption{(a)-(b) Semi-analytical Stokes spectral ratio of the total model $S_\mathrm{R}(\omega_L-\omega_v)$, using formula~\ref{SS}, compared to the standard optomechanical model, $\mathrm{log}\left[S^\mathrm{theo}_\mathrm{R,S}/S^\mathrm{om,theo}_\mathrm{R,S}\right]$  (panel a), and compared to the bare resonant one, $\mathrm{log}\left[S^\mathrm{theo}_\mathrm{R,S}/S^\mathrm{res,theo}_\mathrm{R,S}\right]$ (panel b). (c)-(d) Analogous full-computational Stokes Raman ratios, $\mathrm{log}\left[S_\mathrm{S}/S^\mathrm{om}_\mathrm{S}\right]$ (panel c), and  $\mathrm{log}\left[S_\mathrm{S}/S^\mathrm{res}_\mathrm{S}\right]$ (panel d), respectively. The rest of the parameters coincide with Fig.~1. The convergence for the spectral simulations in the main text and the supplemental material generally requires a number of Fock states truncated to 10 and 3 for the photonic and  vibrational degrees of freedom, respectively. }
	\label{CooperativeStokes}
	\vspace{-0.2cm}
\end{figure}

	Within the manuscript, the study primarily centers around the anti-Stokes peak amplitude, although exploring the behavior at the Stokes Raman scattering frequency is also of interest. Figure~\ref{CooperativeStokes} displays the Stokes amplitude ratio, mirroring the calculation in Figure~3 for the anti-Stokes case. A strong agreement between numerical results and the semi-analytical model is observed, revealing a suppression pattern described by the squared module $\left|C^\mathrm{off}_{r_{S}}+C^\mathrm{res}_{r_{S}}\right|^2$. Unlike the anti-Stokes case, there is no observed enhancement, as $|C^v_{r_{S}}\left(\langle v^\dagger v\rangle,\omega\right)|^2\propto\left(1+\mean{v^\dagger v}\right)$ and, for the considered parameters, $\mean{v^\dagger v}\ll1$.
\section{Spectral behaviour at room temperature}
The behaviour of our system at room temperature is of  relevance for common SERS experimental conditions. For this purpose, we now account for the effect of surrounding vibron thermalization in the standard master equation, being, $\dot{\hat\rho}=-i[\hat H,\hat\rho]+\sum_i\frac{\gamma_i}{2}\mathcal{L}(\hat\xi_i)+\frac{\kappa}{2}\mathcal{L}(\hat a)+\frac{\kappa_v}{2}(n_\mathrm{th}+1)\mathcal{L}(\hat v)+\frac{\kappa_v}{2}n_\mathrm{th}\mathcal{L}(\hat v^\dagger)$, where the vibronic mode thermal occupancy is given by $n_\mathrm{th}=\frac{1}{e^{\hbar \omega_v/K_\mathrm{B}T}-1}$.
\begin{figure}[h]
	
	\begin{center}
		\includegraphics[width=15.2cm]{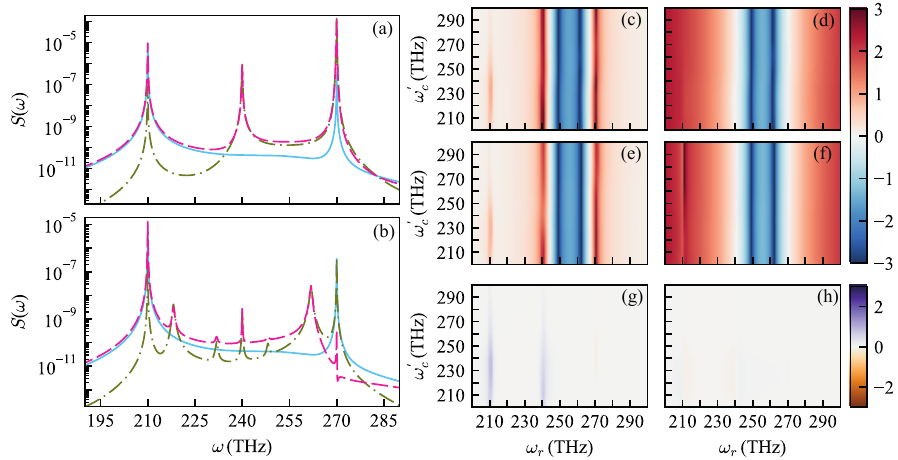}
	\end{center}
	\caption{(a)-(b) Emission spectra $S(\omega)$ for the optomechanical Hamiltonian $H_\mathrm{om}$~\cite{Roelli2016} (solid blue), the resonant SERS Hamiltonian $H_\mathrm{res}$~\cite{Neuman2019} (dashdot olive green), and the total Hamiltonian $H_{\mathrm{om+ res}}$ containing both contributions (dashed magenta) for $T=300$K. Panel (a) describes the enhancement scenario of the anti-Stokes peak observed at $T=0$K with the resonant transition at $\omega_r=270~\mathrm{THz}$, whereas (b) describes a case of suppression with the zero-phonon line (ZPL) at $\omega_r=262~\mathrm{THz}$. The plasmon detuning is set at $\Delta_c^\prime=15~\mathrm{THz}$, whereas $g_\mathrm{om}=0.1~\mathrm{THz}$ and $N=1000$. (c)-(d) Semi-analytical anti-Stokes spectral ratio of the total model $S_\mathrm{R}(\omega_L+\omega_v)$, using formula~(5), compared to the standard optomechanical model, $\mathrm{log}\left[S^\mathrm{theo}_\mathrm{R,aS}/S^\mathrm{om,theo}_\mathrm{R,aS}\right]$  (panel c), and compared to the bare resonant one, $\mathrm{log}\left[S^\mathrm{theo}_\mathrm{R,aS}/S^\mathrm{res,theo}_\mathrm{R,aS}\right]$ (panel d). (e)-(f) Analogous full-computational anti-Stokes Raman ratios, $\mathrm{log}\left[S_\mathrm{aS}/S^\mathrm{om}_\mathrm{aS}\right]$ (panel e), and  $\mathrm{log}\left[S_\mathrm{aS}/S^\mathrm{res}_\mathrm{aS}\right]$ (panel f), respectively. (g) Vibron population ratio comparing the total model with the optomechanical one, $\mathrm{log}\left[\langle v^\dagger v\rangle/\langle v^\dagger v\rangle_\mathrm{om}\right]$, and (h) compared to the bare resonant model, $\mathrm{log}\left[\langle v^\dagger v\rangle/\langle v^\dagger v\rangle_\mathrm{res}\right]$.  The rest of the parameters coincide with Fig.~1 (c).} 
	\label{RoomTemperature}
\end{figure}
As opposed to the results for $T=0$ K in the main text, in Fig.~\ref{RoomTemperature} (a) we do not observe Raman enhancement in the total model but a near-resonant Raman contribution dominance due to both, the high $C^\mathrm{res}_{\mathrm{aS}}$ value around the Raman anti-Stokes line frequency, and the predominance of the thermal population over the optically induced one, which practically equalizes the vibrational population for all the models as we can see in Fig.~\ref{RoomTemperature} (g)-(h). For the Raman suppression conditions, the blue-detuned resonance interference pattern remains similar to the $T=0$ K case (see panels (c)-(d)-(e)-(f)), since $\left|C^\mathrm{off}_{r_{aS}}+C^\mathrm{res}_{r_{aS}}\right|^2$ does not depend on temperature, and so, we observe the same suppression effect in panel (b) as in the results for $T=0$ K.
Increasing the Rabi frequency $\Omega$ powers optomechanical Stokes pumping, potentially surpassing thermal population levels and leading to enhancement effects. It is possible to compute the temperature $T_\mathrm{E}$ at which both vibronic occupancy contributions, thermal and optomechanical, are equal. By performing $n_\mathrm{th}(T_\mathrm{E})=n_\mathrm{f}(\Omega)$, we obtain a fully analytical expression $T_\mathrm{E}=\hbar\omega_v/[K_\mathrm{B}\mathrm{Log}(1+1/n_\mathrm{f}(\Omega))]$, where $n_\mathrm{f}$ denotes the analytical vibron occupancy derived within the linearized optomechanical model as presented in~\cite{Wilson-Rae2008}. Figure~\ref{Thermalpopulation} depicts the anti-Stokes amplitude enhancement as a function of Rabi frequency and temperature, employing parameters consistent with Figure~2(a) in the main manuscript. The ratio experiences an ascent below the equalization temperature $T_\mathrm{E}$, signifying the prevalence of optomechanical Stokes pumping over vibronic thermalization.
\begin{figure}[h]
	\begin{center}
		\includegraphics[width=8.6cm]{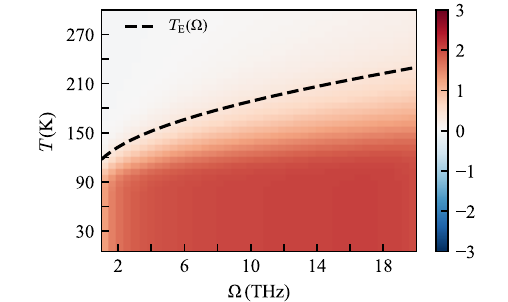}
	\end{center}
	\caption{Full-computational anti-Stokes spectral ratio of the total model $S_\mathrm{R}(\omega_L+\omega_v)$, compared to the bare resonant model, $\mathrm{log}\left[S^\mathrm{theo}_\mathrm{R,aS}/S^\mathrm{res,theo}_\mathrm{R,aS}\right]$ in function of the Rabi frequency $\Omega$ and the temperature $T$. The rest of the parameters coincide with Fig.~2 (a). In black dashed, full-analytical equalization temperature $T_\mathrm{E}$ in function of the rabi frequency $\Omega$.}
	\label{Thermalpopulation}
\end{figure}
Although room temperature enhancement is, as previously mentioned, unattainable, a similar behavior to the $T=0$K case is expected across a wide temperature range.
\begin{figure}[H]
	\begin{center}
		\includegraphics[width=15.2cm]{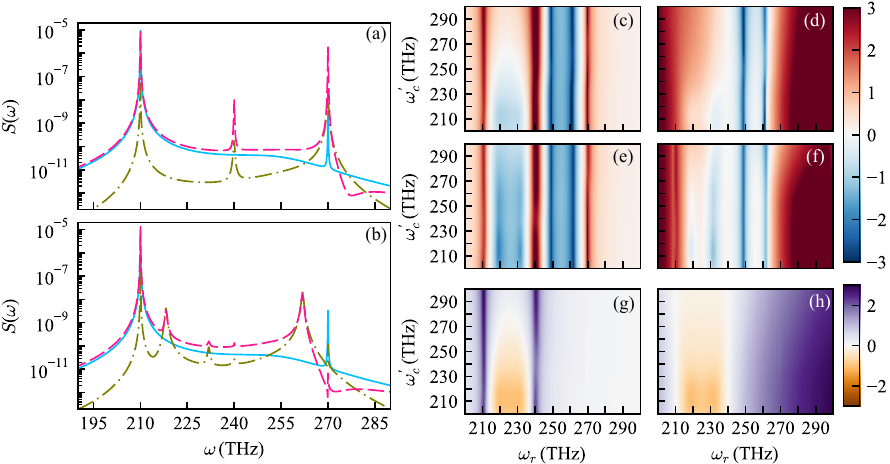}
	\end{center}
	\caption{(a)-(b) Emission spectra $S(\omega)$ for the optomechanical Hamiltonian $H_\mathrm{om}$~\cite{Roelli2016} (solid blue), the resonant SERS Hamiltonian $H_\mathrm{res}$~\cite{Neuman2019} (dashdot olive green), and the total Hamiltonian $H_{\mathrm{om+ res}}$ containing both contributions (dashed magenta) for $T=70$K. Panel (a) describes the enhancement scenario of the anti-Stokes peak observed at $T=0$K with the resonant transition at $\omega_r=270~\mathrm{THz}$, whereas (b) describes a case of suppression with the zero-phonon line (ZPL) at $\omega_r=262~\mathrm{THz}$. The plasmon detuning is set at $\Delta_c^\prime=15~\mathrm{THz}$, whereas $g_\mathrm{om}=0.1~\mathrm{THz}$ and $N=1000$. (c)-(d) Semi-analytical anti-Stokes spectral ratio of the total model $S_\mathrm{R}(\omega_L+\omega_v)$, using formula~(5), compared to the standard optomechanical model, $\mathrm{log}\left[S^\mathrm{theo}_\mathrm{R,aS}/S^\mathrm{om,theo}_\mathrm{R,aS}\right]$  (panel c), and compared to the bare resonant one, $\mathrm{log}\left[S^\mathrm{theo}_\mathrm{R,aS}/S^\mathrm{res,theo}_\mathrm{R,aS}\right]$ (panel d). (e)-(f) Analogous full-computational anti-Stokes Raman ratios, $\mathrm{log}\left[S_\mathrm{aS}/S^\mathrm{om}_\mathrm{aS}\right]$ (panel e), and  $\mathrm{log}\left[S_\mathrm{aS}/S^\mathrm{res}_\mathrm{aS}\right]$ (panel f), respectively. (g) Vibron population ratio comparing the total model with the optomechanical one, $\mathrm{log}\left[\langle v^\dagger v\rangle/\langle v^\dagger v\rangle_\mathrm{om}\right]$, and (h) compared to the bare resonant model, $\mathrm{log}\left[\langle v^\dagger v\rangle/\langle v^\dagger v\rangle_\mathrm{res}\right]$.  The rest of the parameters coincide with Fig.~1 (c).} 
	\label{LNTemperature}
\end{figure}
Figure~\ref{LNTemperature} presents a comparable analysis to that in Figure~\ref{RoomTemperature}, albeit at a temperature slightly below the boiling point of liquid nitrogen, $T=70$K. The outcomes closely resemble those observed in the $T=0$K scenarios depicted in Figures~2 and 3 of the manuscript. This indicates that the enhancement feature of the proposed model persists at more elevated temperatures and is not exclusive to the $T=0$K case.
\section{Stokes-anti-Stokes pair correlations}
\begin{figure}[h]
	\begin{center}
		\includegraphics[width=9cm]{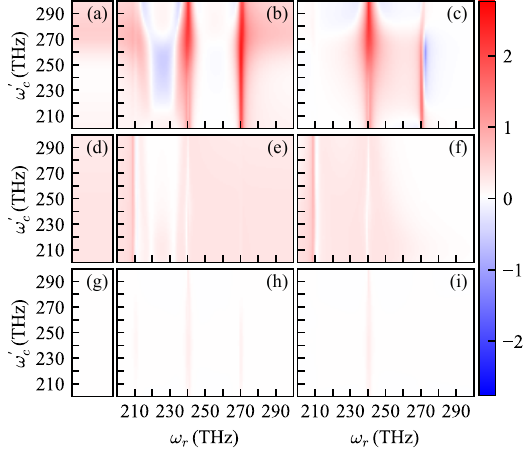}
	\end{center}
	\caption{(a)-(b)-(c) Frequency resolved $g_\Gamma^{(2)}(\omega_\mathrm{aS},\omega_\mathrm{S})$, in logarithmic scale, for the optomechanical Hamiltonian $H_\mathrm{om}$ (panel a), the resonant SERS Hamiltonian $H_\mathrm{res}$ (panel c), and the total Hamiltonian $H_\mathrm{om+res}$ containing both contributions (panel b). Analogous calculations for $g_\Gamma^{(2)}(\omega_\mathrm{S},\omega_\mathrm{S})$ in (d)-(e)-(f) and $g_\Gamma^{(2)}(\omega_\mathrm{aS},\omega_\mathrm{aS})$ in (g)-(h)-(i). The parameters coincide with those in Fig.~4 in the main text.}
	\label{Quantumcorrelations}
\end{figure}

In this section we analyze  the  photon-pair correlations associated to the main cooperative Raman enhancement and suppression processes considered in the manuscript. We employ the sensor method to calculate the frequency-filtered second-order photon correlations $g_\Gamma^{(2)}(\omega_\mathrm{aS},\omega_\mathrm{S})$~\cite{SanchezMunoz2014} for the Stokes and anti-Stokes photons. In Fig.~\ref{Quantumcorrelations} (a)-(c), we observe that the total model (panel b) exhibits a different structure compared to both the optomechanical model (panel a) and the bare near-resonant model (panel c). In particular, the suppression of the anti-Stokes emission depicted in Fig. 2(b) in the main text is associated with a uncorrelated behavior $g_\Gamma^{(2)}(\omega_\mathrm{aS},\omega_\mathrm{S})\approx 1$ in panel (b). Conversely, in the case of the main anti-Stokes enhancement shown in Fig. 2 (a) when the ZPL is near the anti-Stokes frequency, we observe pronounced bunching in comparison to the previous standard optomechanical models, which is associated to the enhanced creation of anti-Stokes photons from Stokes processes. Together with the low values of the single-frequency correlators $g_\Gamma^{(2)}(\omega_\mathrm{S},\omega_\mathrm{S})$ and $g_\Gamma^{(2)}(\omega_\mathrm{aS},\omega_\mathrm{S})$ observed in Fig.~\ref{Quantumcorrelations} (d)-(i), such correlated creation of Stokes-anti-Stokes pairs results in a large violation of the Cauchy-Schwarz violation $R\gg1$ discussed in the main text.

\bibliographystyle{apsrev4-2}
\bibliography{Electronic_dynamics_SERS}